%Paper: cond-mat/9403022
%From: William Putikka <putikka@fangio.magnet.fsu.edu>
%Date: Sun, 6 Mar 94 17:40:01 EST
%Date (revised): Fri, 11 Mar 94 12:41:20 EST

%%%%%%%%%%%%%%%%%%%%%%%%%%%%%%%%%%%%%%%%%%%%%%%%%%%%%%%%%%%%%%%%%%%%%%%
% This is a plain TeX file, including all macros.
% 16 uuencoded, compressed postscript figures are
% appended at the end and should be removed before texing.
%
% Please report problems with printing this paper or requests for
% hard copies of the figures to putikka@theory.nhmfl.fsu.edu.
%
%  Define a whole menagerie of pseudo-12pt fonts
     
\font\twelverm=cmr10 scaled 1200    \font\twelvei=cmmi10 scaled 1200
\font\twelvesy=cmsy10 scaled 1200   \font\twelveex=cmex10 scaled 1200
\font\twelvebf=cmbx10 scaled 1200   \font\twelvesl=cmsl10 scaled 1200
\font\twelvett=cmtt10 scaled 1200   \font\twelveit=cmti10 scaled 1200
     
\skewchar\twelvei='177   \skewchar\twelvesy='60
     
%  Define \...point macros to change fonts and spacings consistently
     
\def\twelvepoint{\normalbaselineskip=12.4pt plus 0.1pt minus 0.1pt
  \abovedisplayskip 12.4pt plus 3pt minus 9pt
  \belowdisplayskip 12.4pt plus 3pt minus 9pt
  \abovedisplayshortskip 0pt plus 3pt
  \belowdisplayshortskip 7.2pt plus 3pt minus 4pt
  \smallskipamount=3.6pt plus1.2pt minus1.2pt
  \medskipamount=7.2pt plus2.4pt minus2.4pt
  \bigskipamount=14.4pt plus4.8pt minus4.8pt
  \def\rm{\fam0\twelverm}          \def\it{\fam\itfam\twelveit}%
  \def\sl{\fam\slfam\twelvesl}     \def\bf{\fam\bffam\twelvebf}%
  \def\mit{\fam 1}                 \def\cal{\fam 2}%
  \def\tt{\twelvett}
  \textfont0=\twelverm   \scriptfont0=\tenrm   \scriptscriptfont0=\sevenrm
  \textfont1=\twelvei    \scriptfont1=\teni    \scriptscriptfont1=\seveni
  \textfont2=\twelvesy   \scriptfont2=\tensy   \scriptscriptfont2=\sevensy
  \textfont3=\twelveex   \scriptfont3=\twelveex  \scriptscriptfont3=\twelveex
  \textfont\itfam=\twelveit
  \textfont\slfam=\twelvesl
  \textfont\bffam=\twelvebf \scriptfont\bffam=\tenbf
  \scriptscriptfont\bffam=\sevenbf
  \normalbaselines\rm}
     
%	tenpoint

%%
%%	Various internal macros
%%
     
\def\beginlinemode{\endmode
  \begingroup\parskip=0pt \obeylines\def\\{\par}\def\endmode{\par\endgroup}}
\def\beginparmode{\endmode
  \begingroup \def\endmode{\par\endgroup}}
\let\endmode=\par
{\obeylines\gdef\
{}}
\def\singlespace{\baselineskip=\normalbaselineskip}

\def\oneandahalfspace{\baselineskip=\normalbaselineskip
  \multiply\baselineskip by 3 \divide\baselineskip by 2}
\def\doublespace{\baselineskip=\normalbaselineskip \multiply\baselineskip by 2}

\newcount\firstpageno
\firstpageno=2
\footline={\ifnum\pageno<\firstpageno{\hfil}\else{\hfil\twelverm\folio\hfil}\fi}
\def\toppageno{\global\footline={\hfil}\global\headline
  ={\ifnum\pageno<\firstpageno{\hfil}\else{\hfil\twelverm\folio\hfil}\fi}}
\let\rawfootnote=\footnote		% We must set the footnote style
\def\footnote#1#2{{\rm\doublespace\parindent=0pt\parskip=0pt
  \rawfootnote{#1}{#2\hfill\vrule height 0pt depth 6pt width 0pt}}}
\def\raggedcenter{\leftskip=4em plus 12em \rightskip=\leftskip
  \parindent=0pt \parfillskip=0pt \spaceskip=.3333em \xspaceskip=.5em
  \pretolerance=9999 \tolerance=9999
  \hyphenpenalty=9999 \exhyphenpenalty=9999 }
\def\dateline{\rightline{\ifcase\month\or
  January\or February\or March\or April\or May\or June\or
  July\or August\or September\or October\or November\or December\fi
  \space\number\year}}
\def\today{\ifcase\month\or
  January\or February\or March\or April\or May\or June\or
  July\or August\or September\or October\or November\or December\fi
  \space\number\day, \number\year}
\def\received{\vskip 3pt plus 0.2fill
 \centerline{\sl (Received\space\ifcase\month\or
  January\or February\or March\or April\or May\or June\or
  July\or August\or September\or October\or November\or December\fi
  \qquad, \number\year)}}
     
%%
%%	Page layout, margins, font and spacing (feel free to change)
%%
     
\hsize=6.5truein
\hoffset=0truein
\vsize=8.9truein
\voffset=0truein
\parskip=\medskipamount
\def\\{\cr}
\twelvepoint		% selects twelvepoint fonts (cf. \tenpoint)
\doublespace		% selects double spacing for main part of paper (cf.
			%	\singlespace, \oneandahalfspace)
\overfullrule=0pt	% delete the nasty little black boxes for overfull box
     
%%
%%	The user definitions for major parts of a paper (feel free to change)
%%
     
%\def\draft{
%\input timestamp\rightline{Draft \timestamp}}  %  "Draft", Timestamp

	% Preprint number at upper right of title page
     
\def\title			%  Title on title page
  {\null\vskip 3pt plus 0.2fill
   \beginlinemode \doublespace \raggedcenter \bf}
     
\def\author			%  Author(s) name(s)  on title page
  {\vskip 3pt plus 0.2fill \beginlinemode
   \singlespace \raggedcenter}
     
\def\affil			% Affiliations (can intermix with \author)
  {\vskip 3pt plus 0.1fill \beginlinemode
   \oneandahalfspace \raggedcenter \sl}
     
\def\abstract			% Begin abstract
  {\vskip 3pt plus 0.3fill \beginparmode
   \doublespace
   \centerline{ABSTRACT} }
     
\def\endtopmatter		% End title page, begin body of paper
  {\endpage			% 	This subsumes \body
   \body}
     
\def\body			% Begin text body;  can be used to end
  {\beginparmode}		% \title, \author, \affil, \abstract,
				% \reference, or \figurecaption modes
     
\def\head#1{			% Head;  NOTE enclose the text in {}
  \goodbreak\vskip 0.5truein	%  e.g., \head{I. Introduction}
  {\immediate\write16{#1}
   \raggedcenter \uppercase{#1}\par}
   \nobreak\vskip 0.25truein\nobreak}

\def\beneathrel#1\under#2{\mathrel{\mathop{#2}\limits_{#1}}}
     
\def\refto#1{$^{#1}$}		% For references in text as superscript
     
\def\references			% Begin references -- basic format is Phys Rev
  {\head{References}		% I.e., volume, page, year (space after commas).
   \beginparmode
   \frenchspacing \parindent=0pt \leftskip=1truecm
   \parskip=8pt plus 3pt \everypar{\hangindent=\parindent}}

\gdef\refis#1{\item{#1.\ }}			% Ref list numbers.
     
\gdef\journal#1, #2, #3, 1#4#5#6{		% Journal reference.  Comma sets
    {\sl #1~}{\bf #2}, #3 (1#4#5#6)}		% off: name, vol, page, year

\def\prb{\journal Phys. Rev. B, }

\def\endreferences{\body}
     
\def\figurecaptions		% Begin figure captions
  {\endpage
   \beginparmode
   \head{Figure Captions}
}

\def\endpage			%  Eject a page
  {\vfill\eject}
     
\def\endpaper			%  Ways to say goodbye
  {\endmode\vfill\supereject}

\def\tag#1$${\eqno(#1)$$}
     
\def\align#1$${\eqalign{#1}$$}

\def\aligntag#1$${\gdef\tag##1\\{&(##1)\cr}\eqalignno{#1\\}$$
  \gdef\tag##1$${\eqno(##1)$$}}
     
\def\endaligntag{}

\def\overset#1\to#2{{\mathop{#2}^{#1}}}
\def\underset#1\to#2{{\mathop{#2}_{#1}}}

%%
%%	Various little user definitions
%%
     
\def\ref#1{Ref.~#1}			% 	for inline references
\def\Ref#1{Ref.~#1}			% 	ditto
\def\[#1]{[\cite{#1}]}
\def\cite#1{{#1}}
\def\(#1){(\call{#1})}
\def\call#1{{#1}}
\def\taghead#1{}
\def\frac#1#2{{#1 \over #2}}

\def\12{{1\over2}}
\def\eg{{\it e.g.,\ }}

\def\etal{{\it et al.\ }}

\def\sla{\raise.15ex\hbox{$/$}\kern-.57em}
\def\leaderfill{\leaders\hbox to 1em{\hss.\hss}\hfill}
\def\twiddle{\lower.9ex\rlap{$\kern-.1em\scriptstyle\sim$}}
\def\bigtwiddle{\lower1.ex\rlap{$\sim$}}
\def\gtwid{\mathrel{\raise.3ex\hbox{$>$\kern-.75em\lower1ex\hbox{$\sim$}}}}
\def\ltwid{\mathrel{\raise.3ex\hbox{$<$\kern-.75em\lower1ex\hbox{$\sim$}}}}
\def\square{\kern1pt\vbox{\hrule height 1.2pt\hbox{\vrule width 1.2pt\hskip 3pt
   \vbox{\vskip 6pt}\hskip 3pt\vrule width 0.6pt}\hrule height 0.6pt}\kern1pt}
\def\tdot#1{\mathord{\mathop{#1}\limits^{\kern2pt\ldots}}}

\def\pmb#1{\setbox0=\hbox{#1}%
  \kern-.025em\copy0\kern-\wd0
  \kern  .05em\copy0\kern-\wd0
  \kern-.025em\raise.0433em\box0 }

\catcode`@=11
\newcount\tagnumber\tagnumber=0
     
\immediate\newwrite\eqnfile
\newif\if@qnfile\@qnfilefalse
\def\write@qn#1{}
\def\writenew@qn#1{}
\def\w@rnwrite#1{\write@qn{#1}\message{#1}}
\def\@rrwrite#1{\write@qn{#1}\errmessage{#1}}
     
\def\taghead#1{\gdef\t@ghead{#1}\global\tagnumber=0}
\def\t@ghead{}
     
\expandafter\def\csname @qnnum-3\endcsname
  {{\t@ghead\advance\tagnumber by -3\relax\number\tagnumber}}
\expandafter\def\csname @qnnum-2\endcsname
  {{\t@ghead\advance\tagnumber by -2\relax\number\tagnumber}}
\expandafter\def\csname @qnnum-1\endcsname
  {{\t@ghead\advance\tagnumber by -1\relax\number\tagnumber}}
\expandafter\def\csname @qnnum0\endcsname
  {\t@ghead\number\tagnumber}
\expandafter\def\csname @qnnum+1\endcsname
  {{\t@ghead\advance\tagnumber by 1\relax\number\tagnumber}}
\expandafter\def\csname @qnnum+2\endcsname
  {{\t@ghead\advance\tagnumber by 2\relax\number\tagnumber}}
\expandafter\def\csname @qnnum+3\endcsname
  {{\t@ghead\advance\tagnumber by 3\relax\number\tagnumber}}
     
\def\equationfile{%
  \@qnfiletrue\immediate\openout\eqnfile=\jobname.eqn%
  \def\write@qn##1{\if@qnfile\immediate\write\eqnfile{##1}\fi}
  \def\writenew@qn##1{\if@qnfile\immediate\write\eqnfile
    {\noexpand\tag{##1} = (\t@ghead\number\tagnumber)}\fi}
}
     
\def\callall#1{\xdef#1##1{#1{\noexpand\call{##1}}}}
\def\call#1{\each@rg\callr@nge{#1}}
     
\def\each@rg#1#2{{\let\thecsname=#1\expandafter\first@rg#2,\end,}}
\def\first@rg#1,{\thecsname{#1}\apply@rg}
\def\apply@rg#1,{\ifx\end#1\let\next=\relax%
\else,\thecsname{#1}\let\next=\apply@rg\fi\next}
     
\def\callr@nge#1{\calldor@nge#1-\end-}
\def\callr@ngeat#1\end-{#1}
\def\calldor@nge#1-#2-{\ifx\end#2\@qneatspace#1 %
  \else\calll@@p{#1}{#2}\callr@ngeat\fi}
\def\calll@@p#1#2{\ifnum#1>#2{\@rrwrite{Equation range #1-#2\space is bad.}
\errhelp{If you call a series of equations by the notation M-N, then M and
N must be integers, and N must be greater than or equal to M.}}\else%
 {\count0=#1\count1=#2\advance\count1 by1\relax\expandafter\@qncall\the\count0,%
  \loop\advance\count0 by1\relax%
    \ifnum\count0<\count1,\expandafter\@qncall\the\count0,%
  \repeat}\fi}
     
\def\@qneatspace#1#2 {\@qncall#1#2,}
\def\@qncall#1,{\ifunc@lled{#1}{\def\next{#1}\ifx\next\empty\else
  \w@rnwrite{Equation number \noexpand\(>>#1<<) has not been defined yet.}
  >>#1<<\fi}\else\csname @qnnum#1\endcsname\fi}
     
\let\eqnono=\eqno
\def\eqno(#1){\tag#1}
\def\tag#1$${\eqnono(\displayt@g#1 )$$}
     
\def\aligntag#1\endaligntag
  $${\gdef\tag##1\\{&(##1 )\cr}\eqalignno{#1\\}$$
  \gdef\tag##1$${\eqnono(\displayt@g##1 )$$}}

\def\eqalignno#1{\displ@y \tabskip\centering
  \halign to\displaywidth{\hfil$\displaystyle{##}$\tabskip\z@skip
    &$\displaystyle{{}##}$\hfil\tabskip\centering
    &\llap{$\displayt@gpar##$}\tabskip\z@skip\crcr
    #1\crcr}}
     
\def\displayt@gpar(#1){(\displayt@g#1 )}
     
\def\displayt@g#1 {\rm\ifunc@lled{#1}\global\advance\tagnumber by1
        {\def\next{#1}\ifx\next\empty\else\expandafter
        \xdef\csname @qnnum#1\endcsname{\t@ghead\number\tagnumber}\fi}%
  \writenew@qn{#1}\t@ghead\number\tagnumber\else
        {\edef\next{\t@ghead\number\tagnumber}%
        \expandafter\ifx\csname @qnnum#1\endcsname\next\else
        \w@rnwrite{Equation \noexpand\tag{#1} is a duplicate number.}\fi}%
  \csname @qnnum#1\endcsname\fi}
     
\def\ifunc@lled#1{\expandafter\ifx\csname @qnnum#1\endcsname\relax}
     
\let\@qnend=\end\gdef\end{\if@qnfile
\immediate\write16{Equation numbers written on []\jobname.EQN.}\fi\@qnend}
     
\catcode`@=12
\catcode`@=11
\newcount\r@fcount \r@fcount=0
\newcount\r@fcurr
\immediate\newwrite\reffile
\newif\ifr@ffile\r@ffilefalse
\def\w@rnwrite#1{\ifr@ffile\immediate\write\reffile{#1}\fi\message{#1}}
     
\def\writer@f#1>>{}
\def\referencefile{%			  Stuff to write .REF file
  \r@ffiletrue\immediate\openout\reffile=\jobname.ref%
  \def\writer@f##1>>{\ifr@ffile\immediate\write\reffile%
    {\noexpand\refis{##1} = \csname r@fnum##1\endcsname = %
     \expandafter\expandafter\expandafter\strip@t\expandafter%
     \meaning\csname r@ftext\csname r@fnum##1\endcsname\endcsname}\fi}%
  \def\strip@t##1>>{}}

\def\citeall#1{\xdef#1##1{#1{\noexpand\cite{##1}}}}
\def\cite#1{\each@rg\citer@nge{#1}}	% Variable No. of args, separated by ","
     
\def\each@rg#1#2{{\let\thecsname=#1\expandafter\first@rg#2,\end,}}
\def\first@rg#1,{\thecsname{#1}\apply@rg}	% each@ag is a general purpose
\def\apply@rg#1,{\ifx\end#1\let\next=\relax%	  variable no. of arg. macro.
\else,\thecsname{#1}\let\next=\apply@rg\fi\next}% args separated by commas
     
\def\citer@nge#1{\citedor@nge#1-\end-}	% Check for M-N range (M and N numbers)
\def\citer@ngeat#1\end-{#1}
\def\citedor@nge#1-#2-{\ifx\end#2\r@featspace#1 % Single argument
  \else\citel@@p{#1}{#2}\citer@ngeat\fi}	% M-N range of arguments
\def\citel@@p#1#2{\ifnum#1>#2{\errmessage{Reference range #1-#2\space is bad.}
    \errhelp{If you cite a series of references by the notation M-N, then M and
    N must be integers, and N must be greater than or equal to M.}}\else%
 {\count0=#1\count1=#2\advance\count1 by1\relax\expandafter\r@fcite\the\count0,%
  \loop\advance\count0 by1\relax%	  Loop from M to N
    \ifnum\count0<\count1,\expandafter\r@fcite\the\count0,%
  \repeat}\fi}
     
\def\r@featspace#1#2 {\r@fcite#1#2,}	% Eat spaces at beginning or end of arg
\def\r@fcite#1,{\ifuncit@d{#1}		% Cite individual reference
    \expandafter\gdef\csname r@ftext\number\r@fcount\endcsname%
    {\message{Reference #1 to be supplied.}\writer@f#1>>#1 to be supplied.\par
     }\fi%
  \csname r@fnum#1\endcsname}
     
\def\ifuncit@d#1{\expandafter\ifx\csname r@fnum#1\endcsname\relax%
\global\advance\r@fcount by1%
\expandafter\xdef\csname r@fnum#1\endcsname{\number\r@fcount}}
     
\let\r@fis=\refis			% Save old \refis, redefine
\def\refis#1#2#3\par{\ifuncit@d{#1}%      Use two params #2 #3 to strip blank
    \w@rnwrite{Reference #1=\number\r@fcount\space is not cited up to now.}\fi%
  \expandafter\gdef\csname r@ftext\csname r@fnum#1\endcsname\endcsname%
  {\writer@f#1>>#2#3\par}}
     
\def\r@ferr{\endreferences\errmessage{I was expecting to see
\noexpand\endreferences before now;  I have inserted it here.}}
\let\r@ferences=\references
\def\references{\r@ferences\def\endmode{\r@ferr\par\endgroup}}
     
\let\endr@ferences=\endreferences
\def\endreferences{\r@fcurr=0%		  Save old \endreferences, redefine
  {\loop\ifnum\r@fcurr<\r@fcount%	  Loop over refnum and produce text
    \advance\r@fcurr by 1\relax\expandafter\r@fis\expandafter{\number\r@fcurr}%
    \csname r@ftext\number\r@fcurr\endcsname%
  \repeat}\gdef\r@ferr{}\endr@ferences}
     
% Save old \endpaper, redefine it to write parting message.
     
\let\r@fend=\endpaper\gdef\endpaper{\ifr@ffile
\immediate\write16{Cross References written on []\jobname.REF.}\fi\r@fend}
     
\catcode`@=12
     
\citeall\refto		% These macros will generate citations
\citeall\ref		%
\citeall\Ref		%

\def\vec#1{{\bf #1}}

\def\dxy{d_{x^2-y^2}}
\def\re{{\rm Re}}
\def\im{{\rm Im}}

\def\sidehead#1{
 \vskip 0.15truecm
 {\noindent #1\par}
 \nobreak\vskip 0.1truein\nobreak}
\def\endtitlepage
 {\endpage
  \body}
\tolerance=2000
\hbadness=2000
\vbadness=2000
\twelvepoint
\centerline{\bf  d-wave Model for Microwave Response of High-$T_c$ 
Superconductors}
\vskip 1cm
\centerline{P.J. Hirschfeld } 
\centerline{{\it Physics Department, University of Florida,}} 
\centerline{{\it Gainesville, FL 32611}}
\vskip .4cm
\centerline{W.O.~Putikka}
\centerline{\it National High Magnetic Field Laboratory, Florida State 
University,}
\centerline{\it Tallahassee, FL 32306}
\vskip .4cm
\centerline{D.J.~Scalapino}
\centerline{\it Physics Department, University of California,}
\centerline{\it Santa Barbara, CA 93106--9530}
\vskip 2.6cm
{\centerline {\bf Abstract}}
\vskip 1cm
We develop a simple theory of the electromagnetic response of a d- wave
superconductor in the presence of potential scatterers of arbitrary
s-wave scattering strength and inelastic scattering by antiferromagnetic
spin fluctuations.  In the clean London limit, the conductivity of such
a system may be expressed in "Drude" form, in terms of a
frequency-averaged relaxation time.  We compare predictions of the
theory with recent data on YBCO and BSSCO crystals and on YBCO films.
While fits to penetration depth measurements are promising, the low
temperature behavior of the measured microwave conductivity appears to
be in disagreement with our results.  We discuss implications for d-wave
pairing scenarios in the cuprate superconductors.

\endtitlepage
\doublespace
\sidehead{\bf I. Introduction}

A remarkable series of recent microwave experiments on high quality
single crystals of
YBCO\refto{Hardypendepth,Hardy1,Hardy2,Hardy3,Bonnhardybigpaper} has
been taken as evidence for d-wave pairing in the high-$T_c$ oxide
superconductors, complementing NMR,\refto{NMR}
photoemission,\refto{photo} and SQUID phase coherence data\refto{squid}
supporting the same conclusion.\refto{dwavereviews} In particular, there
is thus far no alternate explanation for the observation of a term
linear in temperature in the YBCO penetration
depth,\refto{Hardypendepth} other than an unconventional order parameter
with lines of nodes on the Fermi surface.  Several initial questions
regarding discrepancies between this result and previous similar
measurements, which reported a quadratic variation in temperature, have
been plausibly addressed by analyses of the effect of disorder, which
have suggested that strong scattering by defects in the dirtier samples
can account for these differences.\refto{Carbotte,felds}
\vskip .2cm
We have recently attempted to analyze the dissipative part of the
electromagnetic response, i.e. the microwave conductivity $\sigma$,
within the same model of d-wave superconductivity plus strong elastic
scattering, to check the consistency of this appealingly simple
picture.\refto{HPSPRL} We found that the conductivity could be
represented in a Drude-like form in which the normal quasiparticle fluid
density and an average over an energy dependent quasiparticle lifetime
entered.  For microwave frequencies small compared to the average
relaxation rate, the conductivity was found to vary as $T^2$ at low
temperatures approaching $ne^2/\pi\Delta_0m$ at zero temperature.  Here
$\Delta_0$ is the gap maximum over the Fermi surface.  At higher
microwave frequencies, the interplay between the microwave frequency and
the quasiparticle lifetime was found to lead to a nearly linear $T$
dependence over a range of temperatures.  While some of the qualitative
predictions of this model are in agreement with experiment, the
low-temperature $T^2$ predictions for the low-frequency microwave
conductivity differ from the linear-T dependence reported.
\vskip .2cm
The main purpose of this paper is to explore further the overall
consistency of the $d$-wave pairing plus resonant scattering model
predictions for the low-temperature behavior of the electromagnetic
response of the superconducting state.  We will also examine the
electromagnetic response over a wider temperature regime by
phenomenologically including the effects of inelastic spin-fluctuation
scattering.  In the process we intend to provide the derivations of
results reported in our previous short communication,\refto{HPSPRL} and
address various questions raised by it:

\item{1)}To what
extent can the microwave conductivity in a $\dxy$-wave superconducting
state be thought of in direct analogy to transport in a weakly
interacting fermion gas with a normal quasiparticle fluid density
$n_{qp}(T)$ and a relaxation time $\tau(\omega)$ characteristic of nodal
quasiparticles?

\item{2)} Can the temperature dependence of the microwave
conductivity be used to extract information on the quasiparticle
lifetime?

\item{3)} What is the characteristic low-temperature dependence of the
quasiparticle lifetime for resonant impurity scattering in a $\dxy$
superconductor and how does it affect $\sigma$?

\item{4)} What happens at higher temperatures when inelastic processes
enter?

\item{5)} What happens to $\sigma_1(T,\Omega)$, $\lambda(T,\Omega)$ and
the surface resistance $R_s(T,\Omega)$ at higher microwave frequencies?

\item{6)} To what extent can a model with a $\dxy$ gap plus scattering
describe the observed penetration depth and conductivity of the
cuprates?  Can the response of a $\dxy$-wave state be distinguished from
that of a highly anisotropic $s$-wave state?

\vskip .2cm
The plan of this work is as follows.  In section II, we derive the
expressions necessary for the analysis of the conductivity and
penetration depth of a superconductor in the presence of impurities of
arbitrary strength within BCS theory.  In section III, we examine
several useful limiting cases of these results analytically.  In
Sec.~IV, we introduce a natural definition of the quasiparticle lifetime
which allows the conductivity to be cast in a ``Drude-like'' form with a
temperature dependent carrier concentration $n_{qp}(T)$.  Then we
describe results obtained from a model for inelastic scattering by
antiferromagnetic spin fluctuations and include these in a
phenomenological way so as to describe the conductivity over a wider
temperature regime.  In section V, we compare results for the
penetration depth, conductivity and surface impedance with data on
high-quality samples, including both i) scaling tests of the d-wave plus
resonant scattering theory at low temperatures, and ii) fits over the
entire temperature range.  In section VI we present our conclusions
concerning the validity of the model and suggestions for future work.

\sidehead{\bf II. Electromagnetic response: formalism}
      We first review the theory of the current response of a
superconductor with general order parameter $\Delta_k$ to an external
electromagnetic field, with collisions due to elastic impurity
scattering included at the t-matrix level.\refto{Klemm1,PJHcond,PJHskin}
We expect such a theory to be valid at low temperatures in the
superconducting state, if inelastic contributions to the scattering rate
fall off sufficiently rapidly with decreasing temperature.  This is the
case in the model we discuss most thoroughly, namely a $d_{x^2-y^2}$
state with an electronic pairing mechanism.  In such a case, as the gap
opens, the low-frequency spectral weight of the interaction is supressed
and the dynamic quasiparticle scattering decreases.  The scattering rate
in the superconducting state contains two factors of reduced temperature
$T/T_c$ for electron-electron scattering, and one for the available
density of states in the d-wave state, and therefore varies as
$(T/T_c)^3$ at low temperatures.  At temperatures of order $.3-.4T_c$
the dynamic scattering has decreased by one or two orders of magnitude
from its normal state value, at which point elastic impurity scattering
dominates the transport.  In this low temperature region, the gap is
well formed and its frequency dependence occurs on scales larger than
$T_c$.  Thus it is appropriate to model this system within a BCS
framework.  Furthermore, since the dominant quasiparticle density is
associated with the nodal regions, we assume that the qualitative
features of the temperature dependence of the transport will be
unaffected by the details of the band structure, and consider a
cylindrical Fermi surface with density of states $N_0$, and an order
parameter $\Delta_k = \Delta_0 (T){\rm cos} 2 \phi$ confined to within a
BCS cutoff of this surface.  A more complete theory capable of
describing the higher temperature regime where inelastic scattering
processes become important is discussed in section IV.

If an electromagnetic wave of frequency $\Omega$ is normally incident on
a plane superconducting surface, the current response may be written
%\begin{eqnarray}
$$
{\bf j}({\bf q},\Omega)  = 
-{\buildrel\leftrightarrow\over K}({\bf q},\Omega){\bf A}({\bf q},\Omega)
%\nonumber \\
 =-\Big[ {\buildrel\leftrightarrow\over K_p}
({\bf q},\Omega) - {ne^2\over mc}
\Big]{\bf A}({\bf q},\Omega),\eqno()
$$
%\end{eqnarray}
where ${\bf A} $ is the applied vector potential.  The response function
is related simply to the retarded current-current correlation function,
with
%\begin{eqnarray}
$$
\eqalign{ {\buildrel\leftrightarrow\over K}_p({\bf q},\Omega) =
<[{\bf j},{\bf j}]^R>({\bf q},\Omega) 
\simeq\hskip 7cm\cr
\simeq ({-2ne^2\over mc})<\hat k\hat k\int d\xi_k
T\sum_n tr\Big[\underline g({\bf k_+},\omega_n)\underline g
({\bf k_-},\omega_n-\Omega_m)\Big]>_{\hat 
k}\mid_{i\Omega_m\rightarrow\Omega +i0^+} ,}\eqno()
$$
%\end{eqnarray}
where ${\bf k_\pm} \equiv {\bf k\pm} {\bf q}/2$ and $\omega_n=(2n+1)\pi
T$ and $\Omega_m=2m\pi T$ are the usual Matsubara frequencies.  The
approximate equality in the last step above corresponds to the neglect
of vertex corrections due to impurity scattering and order parameter
collective modes.  The former vanish identically at $q=0$ for a singlet
gap and s-wave impurity scattering,\refto{PJHconsequences} while the
latter are irrelevant if the order parameter corresponds to a
nondegenerate representation of the point group.  As usual, in the last
step we have performed the analytical continuation $i\Omega_m\rightarrow
\Omega +i0^+$.  The single particle matrix propagator $\underline g$ is
given as, \eg\ in \ref{PJHconsequences} in terms of its components in
particle-hole space
%\begin{equation}
$$
\underline g(\vec k,\omega_n) = -{i\tilde\omega_n\underline\tau^0 + 
\tilde\xi_k
\underline\tau^3 + \tilde\Delta_k\tau^1\over \tilde\omega_n^2 + 
\tilde\xi_k^2
+\mid{\tilde\Delta}_k\mid^2}\eqno()
$$
%\end{equation}
where the $\underline\tau^i$ are the Pauli matrices and $\tilde\Delta_k$
is a unitary order parameter in particle-hole and spin space.  The
renormalized quantities are given by $\tilde\omega_n =
\omega_n-\Sigma_0(\omega_n)$, 
$\tilde\xi_k = \xi_k + \Sigma_3(\omega_n)$, and $\tilde\Delta_k =
\Delta_k + \Sigma_1(\omega_n)$, where the self-energy due to s-wave
impurity scattering has been expanded $\underline\Sigma =
\Sigma_i\underline\tau^i$.   The renormalization of the single-particle
energies $\xi_k$ measured relative to the Fermi level is required for
consistency even in the s-wave case, although it is frequently neglected
because in the Born approximation for impurity scattering such
renormalizations amount to a chemical potential shift.  For a
particle-hole symmetric system, these corrections can be important for
arbitrary scattering strengths, but are small in either the weak {\it
or} strong scattering limit.\refto{PJHconsequences,Arfi} We therefore
neglect them in what follows, and postpone discussion of the
particle-hole asymmetric case, where these effects can become large, to
a later work.
\vskip .2cm
A further simplification arises for odd-parity states and certain d-wave
states of current interest, where a reflection or other symmetry of the
order parameter leads to the vanishing of the off-diagonal self-energy
$\Sigma_1$.  In this case, the gap is unrenormalized
(${\tilde\Delta}_k=\Delta_k$), leading to a breakdown of Anderson's
theorem and the insensitivity of the angular (e.g., nodal) {\it
structure} of the gap to pairbreaking effects.
\vskip .2cm
Rather than solve the self-consistent problem in full generality, in
most of what follows, we focus on two cases of special interest: i)
s-wave pairing with weak scattering, for purposes of comparison; and ii)
d-wave pairing without $\Delta_k$ renormalization for weak or resonant
s-wave scattering.  In case i), the self-energies $\Sigma_0=\Gamma_NG_0$
and $\Sigma_1=\Gamma_NG_1$ are the familiar integrated Green's functions
from Abrikosov-Gor'kov theory, where $\Gamma_N$ is the scattering rate
at $T_c$ attributable to impurities alone, and we have defined
$G_\alpha\equiv(i/2\pi N_0)\Sigma_k Tr[\underline\tau^\alpha \underline
g]$.  The Green's function (3) and the self-energies must be calculated
together with the gap equation, $\Delta(k) = T\sum_n\sum_{k^\prime}
V_{kk^\prime} {\rm Tr} (\tau_1 / 2) \underline g (k^\prime,\omega_n)$,
where $V_{kk^\prime}$ is the pair potential. In Secs. II-III, all
calculations are done self-consistently within weak-coupling BCS theory,
which yields $\Delta_0/T_c = 2.14$ for a pure $d_{x^2-y^2}$ state.  When
comparing with experimental data in Secs. IV-V, we adopt larger values
of $\Delta_0/T_c $ of 3 or 4 to simulate strong-coupling corrections.
\vskip .2cm
We now continue the derivation of the response on a level sufficiently
general to subsume both cases i) and ii) above.  If we neglect $\xi_k$
renormalizations, the self-energies are given in a $t$-matrix
approximation by
%\begin{equation}
$$
\Sigma_0={{\Gamma G_0}\over{c^2+{G_1}^2-{G_0}^2}}; 
    ~\Sigma_1={{-\Gamma G_1}\over{c^2+{G_1}^2-{G_0}^2}},\eqno()
$$
%\end{equation}
where $\Gamma\equiv n_i n/(\pi N_0)$ is a scattering rate depending only
on the concentration of defects $n_i$, the electron density $n$, and the
density of states at the Fermi level, $N_0$, while the strength of an
individual scattering event is characterized by the cotangent of the
scattering phase shift, $c$. The Born limit corresponds to $ c \gg 1$,
so that $\Gamma/c^2
\simeq \Gamma_N$, while the unitarity limit corresponds to $c=0$.
To evaluate Eq. (2), we first perform the frequency sums, then perform
the energy integrations as in \ref{PJHskin}, yielding in the general
case
\vskip .2cm
%\begin{eqnarray}
$$
\eqalign{{\rm Re}\,{\buildrel\leftrightarrow\over K}(\vec q,\Omega)
 = {1\over 2}{ne^2\over mc} \int {{d\phi} \over {2 \pi}}  
\hat k : \hat k\int d\omega\Big\{\Big[{\rm tanh}{\beta\omega\over 2}-
{\rm tanh} \beta {(\omega-\Omega)\over 2}\Big]{\rm Re}
\tilde I_{+-}(\omega,\omega - \Omega)+\cr
\qquad +\Big[{\rm tanh} {\beta\omega\over 2}+{\rm tanh}{\beta
(\omega-\Omega)
\over 2}\Big]{\rm Re}\tilde I_{++}
(\omega,\omega - \Omega)\Big\},}\eqno()
$$
%\end{eqnarray}
%\begin{eqnarray}
\vskip .2cm
$$
\eqalign{{\rm Im}\,{\buildrel\leftrightarrow\over K}(\vec q,\Omega) = 
-{1\over 2}
{ne^2\over mc}\int {d\phi\over 2\pi}\hat k:\hat k\int d\omega\Big\{
\Big[{\rm tanh} {\beta\omega\over 2} - {\rm tanh}{\beta(\omega-\Omega)
\over
2}\Big]\times\cr
\quad \times {\rm Im}\{\tilde I_{++}(\omega,\omega-\Omega)
-\tilde I_{+-}(\omega,\omega-\Omega)\}\Big\}}.\eqno()
$$
%\end{eqnarray}
\vskip .2cm
In calculating the surface impedance of the cuprate superconductors, it
is important to take into account the anisotropy of these layered
materials.\refto{DJSq=0} Here we are interested in the response
associated with currents which flow in the $ab$ layers.  The wavevector
in the $ab$ plane is determined by the long wavelength of the microwaves
and hence can be set to zero.  Furthermore, the short quasiparticle mean
free path in the $c$-direction means that the surface impedance is
determined by the conductivity of a CuO$_2$ layer.  Thus the surface
impedance in this case is given by
$$
Z(\Omega,T) = \left( {i4\pi\Omega \over
c^2(\sigma_1(\Omega,T)-i\sigma_2(\Omega,T)} \right)^{1/2}.\eqno()
$$ 
Here $\sigma_1-i\sigma_2$ is the complex frequency- and
temperature-dependent $q=0$ layer conductivity. It is customary to write
the imaginary part of the conductivity in terms of a frequency- and
temperature-dependent inductive skin depth $\lambda(\Omega,T)$,
$$
\sigma_2={c^2\over 4\pi\Omega\lambda^2(\Omega,T)}.\eqno()
$$
At temperatures a few degrees below $T_c$, $\sigma_2 \gg \sigma_1$, so
that the surface resistance $R_s$ is given by
$$
R_s = \re\, Z(\Omega,T) \cong 
{8\pi^2\Omega^2\lambda^3(\Omega,T)\sigma_1(\Omega,T) \over c^4},\eqno()
$$
and the surface reactance $X_s$ is
$$
X_s = \im\, Z(\Omega,T)\cong {4\pi\Omega\lambda(\Omega,T) \over c^2}.
\eqno()
$$ 
Thus microwave surface impedance measurements provide information on
the inductive skin depth $\lambda(\Omega,T)$ and the real part of the
conductivity $\sigma_1(\Omega,T)$.  In the previous section, we have
dropped the subscript 1 and denoted the real part of the conductivity
simply by $\sigma(\Omega,T)$, and in the limit $\Omega\rightarrow 0$,
$\lambda(0,T)$ is just the London penetration depth.

\vskip .2cm
At $q=0$, the energy-integrated bubbles $\tilde I_{++}$ and $\tilde
I_{+-}$ are given by\refto{PJHskin}
\vskip .2cm
%\begin{equation}
$$
\tilde I_{++}(\omega,\omega')= {1\over\xi_{0+}}-
{{{{\tilde\omega}_+}^\prime({{\tilde\omega}_+}+{{\tilde\omega}_+}
^\prime)
+{{\tilde\Delta}_{k+}}^\prime({{\tilde\Delta}_{k+}}-
{{\tilde\Delta}_{k+}}^\prime)
\over{(\xi_{0+}+\xi_{0+}^\prime)\xi_{0+}\xi_{0+}^\prime}}}\eqno()
$$
%\end{equation}
\vskip .2cm
and
\vskip .2cm
%\begin{equation}
$$
\tilde I_{+-}(\omega,\omega')={1\over\xi_{0+}}+
{{{{\tilde\omega}_-}^\prime({{\tilde\omega}_+}+{{\tilde\omega}_-}
^\prime)
+{{\tilde\Delta}_{k-}}^\prime({{\tilde\Delta}_{k+}}-
{{\tilde\Delta}_{k-}}^\prime)
\over{(\xi_{0+}-\xi_{0-}^\prime)\xi_{0+}\xi_{0-}^\prime}}}.\eqno()
$$
\vskip .2cm
%\end{equation}
\noindent
Here $\tilde\omega_\alpha \equiv \tilde\omega(\omega +i\alpha 0^+)$,
$\tilde{\Delta}_{k\alpha} \equiv \tilde{\Delta}_k (\omega+i\alpha
0^+)$, and $\xi_{0\alpha}\equiv {\rm
sgn}~\omega\sqrt{\tilde\omega_\alpha^2 -\tilde{\Delta}_{k\alpha}^2}$
with $\alpha = \pm 1$.
\vskip .2cm
We first consider the dissipative part of the response, reflected in the
$q=0$ conductivity ${\buildrel\leftrightarrow\over\sigma}(\Omega)=
-(c/\Omega) {\rm Im}\, {\buildrel\leftrightarrow\over K} (q=0,\Omega)$.
Combining Eqs.~(6,11-12) yields
\vskip .2cm
%\begin{eqnarray}
$$
\sigma_{ij}(\Omega)=-{{ne^2}\over{2m\Omega}}
{\int_{-\infty}}^\infty
d\omega \{\tanh[{1\over 2} \beta \omega]-\tanh[{1\over 2} \beta
(\omega-\Omega)]\}
\times S_{ij}(\omega,\Omega),\eqno()
$$
%\end{eqnarray}
\vskip .2cm
\noindent
where
%\begin{eqnarray}
\vskip .2cm
$$
\eqalign{S_{ij}(\omega,\Omega)= Im 
\int {{d\phi}\over{2\pi}} {\hat k}_i {\hat k}_j \times
\Biggl[
{{{\tilde \omega_+}^\prime({\tilde \omega_+}+{\tilde \omega_+}^\prime)
+{{\tilde{ \Delta}_{k+}}^\prime({\tilde {\Delta}_{k+}}- {\tilde
{\Delta}_{k+}}^\prime)}
\over{(\xi_{0+}^2-\xi_{0+}^{\prime 2})}}}
\Bigl({1\over{\xi_{0+}}^\prime}-
{1\over{\xi_{0+}}}\Bigr)+\cr
   ~~+{{{\tilde \omega_-}^\prime
({\tilde \omega_+}+{\tilde \omega_-}^\prime)
+{{\tilde{ \Delta}_{k-}}^\prime({\tilde {\Delta}_{k+}}-
{\tilde {\Delta}_{k-}}^\prime)}
\over{(\xi_{0+}^2-\xi_{0-}^{\prime 2})}}}
\Bigl({1\over{\xi_{0+}}}+
{1\over{\xi_{0-}}^\prime}\Bigr) \Biggr],}\eqno()
$$
%\end {eqnarray}
\vskip .2cm
\noindent
and primed quantities are evaluated at $\omega-\Omega$.  For $d$-wave
pairing there is no gap renormalization, so that
$\tilde\Delta_{k\alpha}=\Delta_k$ and the kernel $S_{ij}$ reduces to
\vskip .2cm
%\begin{equation}
$$
S_{ij}(\omega,\Omega)= Im 
\int {{d\phi}\over{2\pi}} {\hat k}_i {\hat k}_j \times
\Biggl[{{\tilde \omega_+}^\prime\over{{\tilde \omega_+}-
{\tilde \omega_+}^\prime}}\Bigl({1\over{\xi_{0+}}^\prime}-
{1\over{\xi_{0+}}}\Bigr)+{{\tilde \omega_-}^\prime\over{{\tilde 
\omega_+}-
{\tilde \omega_-}^\prime}}\Bigl({1\over{\xi_{0+}}}+
{1\over{\xi_{0-}}^\prime}\Bigr) \Biggr].\eqno()
$$
\vskip .2cm
%\end{equation}
We also require an appropriate expression for the London limit Meissner
kernel ${\rm Re}\,{\buildrel\leftrightarrow\over K}(0,0)$ to evaluate
the penetration depth.  Taking $\Omega\rightarrow 0$ in Equation (5), we
obtain\refto{Grossetal,ChoiMuzikar}
\vskip .2cm
$$ 
{\rm Re}\, K_{ij}(0,0) = -{ne^2\over mc} \int d\omega ~ {\rm tanh}\,
{{\beta\omega}\over 2}\,\int {d\phi\over 2\pi}\,
\hat k_i\hat k_j\, 
{\rm Re} \, \Big\{ {{{\tilde\Delta}_k^2}\over {\xi_{0+}^3}}
\Big\}.\eqno()
$$
\vskip .2cm
\noindent
In the special case of isotropic $s$-wave pairing and Born scattering
this reduces to the well known result\refto{AGD,Skalskietal}
$$
{\rm Re}\, K(0,0)
= -{ne^2\over 2mc} \int d\omega ~
{\rm tanh} {{\beta\omega}\over 2}~ {\rm Re}\,
\Big\{
{\Delta^2\over (v^2-\Delta ^2)[\sqrt{v^2-\Delta ^2}+i\Gamma_N]}
\Big\} ,\eqno()
$$
\vskip .2cm
\noindent
with $v={\tilde{\omega}}_+\Delta/\tilde\Delta$.
\vskip 1cm
\sidehead{\bf III. Limiting cases}
%\vskip 1cm
We are primarily interested in the low-temperature, low-frequency
conductivity required to discuss experiments in the microwave regime.
Since the microwave energy is generally lower than the temperatures of
interest, it is useful to replace $(\tanh {{\beta\omega}/2 }- \tanh
{{\beta (\omega-\Omega)}/2} )/(2\Omega)$ by its small $\Omega/T$ limit $
-\partial f/\partial\omega$, providing an exponential cutoff above the
temperature $T$ in the integral (12).  At low temperatures $T \ll
\Delta_0$, the temperature dependence of the conductivity depends
strongly on the lifetime of the low-energy quasiparticle states,
determined by the self-consistent solution to $\tilde
\omega=\omega-\Sigma_0$ and $\tilde \Delta_k=\Delta_k-\Sigma_1$, where
$\Sigma_0$ and $\Sigma_1$ are given by Eq.~(4).
\vskip .2cm
In an ordinary superconductor with weak scattering, only the
exponentially small number of quasiparticles above the gap edge
contribute to absorption.  Resonant scattering, such as occurs in the
case of a Kondo impurity in a superconductor, is known to give rise to
bound states near the Fermi level, reflected in a finite density of
states at $\omega=0$ and leading to absorption below the gap
edge.\refto{MH} A similar phenomenon occurs in unconventional
superconductors, with the difference that, whereas in the $s$-wave
(Kondo) case the bound state ``impurity band'' is isolated from the
quasiparticle density of states above the gap edge, in unconventional
states with nodes the ``bound state'' lies in a continuum, and the
lifetimes of all states are finite.\refto{PJHresonant, SchmittRinketal}.
Nevertheless the energy range between zero and the gap edge $\Delta_0$
may be partitioned crudely into two regimes, separated by a crossover
energy or temperature $T^*$ dependent on the impurity concentration and
phase shift.  Below $\omega\simeq T^*$, the scattering rate $-2Im
\Sigma_0(\omega)$ is large compared to $\omega$, and the effects of
self-consistency are important.  The physics of this regime is similar
to gapless superconductivity as described by the well-known Abrikosov-
Gor'kov\refto{AG} theory of pairbreaking by magnetic impurities in
ordinary superconductors.  The low-temperature thermodynamic and
transport properties are given by expressions similar to analogous
normal state expressions, with the usual Fermi surface density of states
$N_0$ replaced by a residual density of quasiparticle states
$n_0=N(\omega\rightarrow 0)$ in the superconductor.  Above $T^*$,
self-consistency can be neglected, and transport coefficients are
typically given by power laws in temperature reflecting the nodal
structure of the order parameter.\refto{PethickPines} We note that this
"pure" regime will correspond to the entire temperature range if the
impurity concentration is so small that $T^*\rightarrow 0$.
\vskip .2cm
In this paper we focus primarily on the case of resonant scattering in
an attempt to describe the physics of Zn doping in the cuprate
superconductors.  While Zn impurities are believed to have no, or very
small, magnetic moments\refto{Walstedt}, they nevertheless appear to act
as strong pairbreakers.\refto{Walstedt,NMR} A possible explanation for
this strong scattering could be associated with the fact that an inert
site changes the local spin correlations of its nearest and next nearest
neighbors\refto{Bulut3} These changes can lead to strong
scattering\refto{MonthouxPines} and even to bound state
formation\refto{Poil} for the holes of the doped system.  With this in
mind, here we assume that a Zn impurity may be approximated by an
isotropic {\it potential} scatterer with a large phase shift close to
$\pi/2$.
\vskip .2cm
The essential physics of gapless transport in unconventional
superconductors was discussed in the context of heavy fermion
superconductivity by Hirschfeld et al.\refto{PJHresonant} and
Schmitt-Rink et al.\refto{SchmittRinketal} Although both works presented
calculations for model $p$-wave states, most conclusions reached
regarding $p$-wave states with lines of nodes continue to hold for the
$d$-wave states in quasi-two-dimensional materials of interest here. For
example, the normalized density of states $N(\omega)\equiv -{\rm Im}
G_0(\omega)$ is linear in energy for the pure system, and varies as
$n_0+aT^2$ for $T\ll T^*$ for an infinitesimal concentration of
impurities.  Neresesyan et al.\refto{Tsvelick} have recently called into
question the existence of the residual density of states $n_0$ in a
strictly 2D system.  We believe nevertheless that both the underlying
three--dimensional character of the layered cuprates, as well as the
extremely low temperature at which the difference between the
logarithmic term and the slow power law behavior found in
\ref{Tsvelick} becomes significant,  make such considerations 
irrelevant for our purposes.  
\vskip .2cm
All quantities of interest in the gapless regime may be obtained by
expanding $\tilde \omega$ (and $\tilde\Delta_k$ if necessary) for
$\omega \ltwid T^*$, with the result $\tilde \omega
\simeq  i(\gamma+b\omega^2)+a\omega,$ where $\gamma$, $a$, and 
$b$ are constants. $T^*$ itself may be shown to be of order $\gamma$.
In the case of a $d_{x^2-y^2}$ state over a cylindrical Fermi surface,
$\gamma$ satisfies the self-consistency relation $\gamma=\Gamma
n_0/(c^2+{n_0}^2)$, where
%$n_0=(2\gamma/\pi){\bf K}\left(\Delta_0/
%\sqrt{\gamma^2+{\Delta_0}^2}\right)/\sqrt{\gamma^2+{\Delta_0}^2}
$n_0={2/\pi}{\bf K}(i\Delta_0/\gamma)$, with ${\bf K}$ the complete
elliptic integral of the first kind.  For small impurity concentrations
such that $\Gamma\ll \Delta_0$, one finds $n_0\simeq
(2\gamma/\pi\Delta_0)\, \ln(4\Delta_0/\gamma)$.  In the Born limit,
$c\gg1$, $\gamma\simeq \Gamma_N n_0$, and both $\gamma$ and $n_0$
therefore vary as $\sim \Delta_0\exp(-\Delta_0/\Gamma_N)$.  In the
resonant scattering case of primary interest, on the other hand,
$\gamma=\Gamma/n_0$ and for small concentrations the residual scattering
rate is determined by $(\gamma/\Delta_0)^2=(\pi\Gamma)/[2\Delta_0
\ln(4\Delta_0/\gamma)]$.  
The constants $a$ and $b$ are found to be ${1\over 2}$ and
$-1/(8\gamma)$, respectively.  Thus for strong scattering both $\gamma$
and the residual density of states $n_0$ vary as $(\Gamma
\Delta_0)^{1/2}$ up to a logarithmic correction.  This is important
because it means that low-energy states may be strongly modified, even
though the impurity scattering rate, which varies as $\Gamma$ near
$T_c$, is insufficient to suppress $T_c$ significantly.  In the usual
Born limit, on the other hand, gapless effects become important only
when $\Gamma_N\simeq\Delta_0$, implying a large $T_c$ supression.  As
the normal state inelastic scattering rate, of order $T_c$ in
temperature units, is much larger than the impurity scattering rate in
clean samples, we expect that impurities are in any case relatively
ineffective in suppressing $T_c$ until the elastic scattering rate at
the transition becomes a significant fraction of the inelastic one (see
Sections IV and V).

\vskip .2cm
These estimates enable an immediate evaluation of Eqs.~(13) and (15) in
the gapless regime,
$$
%\begin{equation}
\sigma_{xx}(\Omega=0,T)\simeq  
 {\sigma_{00}\Big[1+{{\pi^2}\over{12}}\Bigl
({T\over\gamma}\Bigr)^2\Big]}\eqno()
$$
%\end{equation}
where $\sigma_{00} = ne^2/(m\pi\Delta_0(0))$ for a $d_{x^2-y^2}$ state.
The first term in Eq.~(18) is a remarkable result first pointed out by
P.A.~Lee,\refto{PALee} namely that the residual conductivity
$\sigma(\Omega\rightarrow 0, T\rightarrow 0)$ of an anisotropic
superconductor with line nodes on the Fermi surface is nonzero and {\it
independent of impurity concentration} to leading order.  It arises
technically from the first term on the right hand side of Eq.~(14), and
is present in principle regardless of the scattering strength.
Physically this reflects a cancellation between the impurity-induced
density of states and the impurity quasiparticle scattering lifetime.
The linear variation $\omega/\Delta_0$ of the d-wave density of states
is cut off when $\omega$ drops below the impurity scattering rate
$\tau^{-1}$.  Therefore, at low energies there is a finite
impurity-induced density of states which varies as
$(\Delta_0\tau)^{-1}$.  At low temperatures such that $T<\tau^{-1}$, the
effective relaxation rate which determines the conductivity is
proportional to the density of states $(\Delta_0\tau)^{-1}$ multiplied
by $\tau$, giving $\Delta_0^{-1}$ independent of the scattering
strength.  Very recently it was pointed out that a generalization of the
present theory to include a finite scattering {\it range} results, in
the limit of sufficiently large range or disorder, in a residual
conductivity which scales with the scattering time
$(2\Gamma)^{-1}$.\refto{Balatsky} The predicted residual conductivity in
this regime is however too small to apply to the experiments considered
here.

\vskip .2cm
     In Figures 1 and 2 we illustrate the effect of varying the phase
shift and impurity concentration on the $T$--dependence of the
conductivity with a full self-consistent numerical evaluation of
Eqs.~(13) and (15) for a $d_{x^2-y^2}$ state.  The intrinsic gapless
behavior represented by Eq.~(18) is clearly visible in the resonant
limit, $c\simeq 0$, but in the Born limit, $c\gg 1$, the same limiting
behavior is effectively unobservable for small concentrations at
$\Omega=0$.  Instead, the conductivity tends to a value
$\sigma_0=ne^2/2m\Gamma_N$ except at exponentially small temperatures,
where it again approaches $\sigma_{\scriptscriptstyle 00}$, due to the
narrow width $\gamma\sim\Delta_0
\exp{-\Delta_0/\Gamma_N}$ of the gapless range in this limit.
\vskip .2cm
For $T>T^*\simeq\gamma$, we take $\tilde
\omega-\omega\simeq\Sigma_0(\omega)$ rather than
$\Sigma_0(\tilde\omega)$, and keep only the leading singular terms in
Eqs.~(13) and (15) as $\Gamma\rightarrow 0$, arriving at the remarkably
simple expression,
%\begin{equation}
$$
\sigma_{xx}(\Omega)\simeq\Bigl({{ne^2}\over{m}}\Bigr){{\int_{-
\infty}
}^{\infty}} d\omega \Bigl({{-\partial f}\over{\partial\omega}} \Bigr) 
%\bigl|{{\omega}\over{\Delta_0}}\bigr|
N(\omega)
   ~{\rm Im }\Bigl({1\over{\Omega-i/\tau (\omega ) }}\Bigr),\eqno()
$$
%\end{equation}
where $\tau^{-1}(\omega)=-2{\rm Im}~\Sigma_0(\omega)$, for any choice of
phase shift.  Note that $N(\omega)$ is the density of states for a pure
superconductor normalized to $N(0)$ and varies as $|\omega/\Delta_0|$
for a $d_{x^2-y^2}$ state at low energies.  Eq.~(19) is exactly the
result expected for the conductivity of noninteracting fermions with
density of states $N(\omega)$ and 1--body relaxation time
$\tau(\omega)$, and is reminiscent of the Drude-like expression used by
Bonn \etal\ to analyze their data.  However, as pointed out in
\ref{HPSPRL}, the $\omega$-dependence of the superconducting density of
states tends to induce a strong energy dependence
in $\tau(\omega)$ in either the strong or weak scattering limits.  
For a $d_{x^2-y^2}$ state we find
$$
\tau^{-1}(\omega)\simeq \Bigl\{
{ {(\pi^2\Gamma\Delta_0)/
[2\omega\ln^2(4\Delta_0/\omega)]\hskip 2cm c\simeq 0 }\atop
{    ~~~(4\Gamma_N\omega/\pi\Delta_0)\ln(4\Delta_0/\omega) 
\hskip 2cm c\gg 1}  }\eqno()
$$
leading to the pure limit conductivity result for 
$\Omega\ll\Gamma
\Delta_0/T$, $T\ll T_c$,
%\begin {equation}
$$
\sigma_{xx}(\Omega=0,T)\simeq \Bigg\{
{ {{2\over 3}\sigma_0
({{T}\over{\Delta_0}})^2
\ln^2 {{4\Delta_0}\over {T}}
\hskip 2cm  c\simeq 0}
\atop
{\sigma_0\hskip 4.8cm  c\gg 1} }.\eqno()
$$
%\end{equation}
In the opposite limit $\Omega\gg\Gamma\Delta_0/T$, $T\ll T_c$ we find
%\begin{equation}
$$
\sigma_{xx}(\Omega,T)\simeq\Biggl\{ { {\Bigl({{ne^2}\over{m}}\Bigr)
{{\pi^2\Gamma}\over{2\Omega^2}} \ln^{-2} {{4\Delta_0}\over{T}}
\hskip 2cm c\simeq 0}\atop
{\Bigl({{ne^2}\over{m}}\Bigr){{4\pi\Gamma_NT^2}\over
{3\Omega^2\Delta_0}}\ln {{4\Delta_0}\over{T}}\hskip 2cm c\gg 1} }.\eqno()
$$
%\end{equation}
\vskip .2cm
It is instructive to compare the form of the previous results with the
more familiar form of those expected for an s-wave superconductor with
weak potential scattering.  We begin with Eqs.~(13) and (14), and
proceed as before in the pure regime, neglecting self-consistency in
$\Sigma_0$ and $\Sigma_1$.  We find
$$
\sigma_{xx}(\Omega)\simeq\Bigl({{ne^2}\over{m}}\Bigr)2\int_\Delta
^{\infty} d\omega \Bigl({{-\partial f}\over{\partial\omega}} \Bigr) 
N(\omega ) ~ {\rm Im} 
\Bigl({1\over{\Omega-i/\tau (\omega ) }}
\Bigr),~~~~~~\hbox{($s$-wave, Born}
)\eqno()
$$ 
where now however the quasiparticle relaxation time in the $s$-wave
superconducting state is given by
%\end{equation}
$(2\tau)^{-1}=- {\rm Im} \Sigma_0(\omega)-(\Delta/\omega) {\rm
Im}\Sigma_1(\omega)$, and $N(\omega)=\omega/\sqrt{\omega^2-\Delta^2}$.
This relaxation rate has a similar form to that found, \eg, by Kaplan
\etal\refto{Kaplanetal} for the electron-phonon quasiparticle relaxation
in ordinary superconductors.  In the limit $\Omega\rightarrow 0,
T\rightarrow 0$, we find
$$
\sigma_{xx}(\Omega)\simeq {ne^2\over m\Gamma_N}\; {\Delta\over T}\,
e^{-\Delta/T} \ln\left({\Delta\over\Omega}\right), \eqno()
$$ 
which is similar in form to the well-known Mattis and Bardeen
result.\refto{MattisBardeen}\vskip .2 cm The hydrodynamic limit results
Eqs.~(21) predict a $T^2$ behavior\refto{HPSPRL} for resonant scattering
or a constant\refto{Coffey} behavior for weak scattering for the low-T
conductivity of a $d$-wave superconductor under the assumptions set down
above.  Neither of these is consistent with the linear-$T$ variation
reported in experiment, which would correspond to the assumption of a
constant relaxation time $\tau$.  Thus the low-temperature experimental
results appear to be inconsistent with the simplest $d$-wave
model.\refto{HPSPRL} However, different physical relaxation mechanisms
than those considered here could change the low-temperature behavior.
\vskip .2cm
The crossover regime between the hydrodynamic (Eq.~(21)) and
collisionless (Eq.~(22)) limits is an interesting one which we
investigate further here.  In Fig.~2, we illustrate this crossover in
the Born limit for a $\dxy$ gap, demonstrating that the result
$\sigma_{xx}\rightarrow\sigma_0$ holds only in the hydrodynamic regime
$\Omega\ll \Gamma_N$.  This is a point of some importance, since
experiments on $Zn$-doped samples appear to indicate a residual
conductivity $\sigma(T\rightarrow 0)$ which scales inversely with
impurity concentration, reminiscent of the zero-frequency Born result
Eq.~(21).  On the other hand, Figure (2) shows that this behavior
disappears at microwave frequencies comparable to those used in the
experiments.  It therefore appears unlikely to us that an explanation in
terms of weak scattering can be compatible with the observations
reported in \ref{Hardy3} and \ref{Bonnhardybigpaper}.
\vskip .2cm
In Fig.~3, we plot the low-temperature conductivity for the case of
resonant scattering to display the same crossover.  It is interesting to
note that a quasilinear behavior is in fact obtained over an
intermediate range of temperatures when the frequency becomes comparable
to the scattering rate, but this behavior does not appear to hold very
far from $\Omega\simeq\Gamma$.
\vskip .2cm
To close the discussion of the low-energy behavior of the conductivity,
we give analytical results for the frequency-dependent conductivity at
zero temperature.\refto{Putikka1} In this case the factor $(\tanh
{{\beta\omega}/2 }- \tanh {{\beta (\omega-\Omega)}/2} )$ appearing in
Eq.~(13) reduces to a window function limiting the range of integration
from 0 to $\Omega$.  The result may be expanded for small values of the
integration variable, yielding in the resonant limit
$$
\sigma_{xx}\simeq\Bigl\{ 
{ {\sigma_{00}\big[ 1 + {1/24} \Big( 
{\Omega/\gamma}\Big) ^2\log^{-1}(4\Delta_0/\gamma)\big]
\hskip 2cm\Omega\ll\gamma}\atop
{  \Bigl({{ne^2}\over{m}}\Bigr)
{{\pi^2\Gamma}\over{2\Omega^2}} \ln^{-2} 
{{4\Delta_0}\over{\gamma}}\hskip 5cm\Omega\gg\gamma}
}.\eqno()
$$ 
In Fig.~4, we plot the frequency dependence of the $T=0$ conductivity
in the impurity-dominated regime.
\vskip .2cm
     A full analysis of surface impedance measurements requires, in
addition to the conductivity $\sigma$, a knowledge of the inductive skin
depth $\lambda(\Omega,T)$, which reduces in the limit $\Omega\rightarrow
0$ to the usual London penetration depth $\lambda(T)$.  The $\Omega=0$
penetration depth in a $d_{x^2-y^2}$ state in the presence of resonant
impurity scattering has been calculated by several authors.  In the
gapless regime $T<T^*$, the linear-$T$ behavior characteristic of a
$d$-wave system is destroyed, and one finds the result $\lambda \simeq
\tilde\lambda_0 +
\pi\lambda_0T^2/(6\gamma \Delta_0)$, where 
$\lambda_0=\sqrt{mc^2/4\pi ne^2}$ is the pure London depth, and the
renormalized zero-$T$ penetration depth is given by\refto{felds}
$({\tilde\lambda}_0-\lambda_0)/\lambda_0\simeq (\gamma/(\pi\Delta_0))
\ln(4\Delta_0/\gamma)\simeq\Gamma/(2\gamma)$.  At higher temperatures
$T^* \ltwid T \ll T_c$, the penetration depth crosses over to the pure
result, $\lambda(T)\simeq\lambda_0[1+\ln 2(T/\Delta_0)]$.  For
completeness, we show in Figure 5 the increase of the zero-temperature
London penetration depth for large values of the scattering parameters
in the Born and unitary limits.  These results are in agreement with
those of Kim et al.\refto{KimMuzikar}
\vskip .2cm
      The presence of low-energy quasiparticles can induce a strong
frequency dependence to the {\it low-temperature} inductive skin depth
$\lambda(T,\Omega)$, which can in some cases mimic shifts in
low-temperature power laws.  Some of these effects were explored in the
context of heavy fermion superconductivity.\refto{PJHresbogo} Here we
observe that the skin depth temperature dependence can be {\it
suppressed} if the microwave frequency is large enough such that
$\Omega\tau>1$.  In this case, it is necessary to use the penetration
depth measured at $\Omega$ rather than the limiting low frequency
penetration depth, to extract the conductivity from surface resistance
data.  A simple expression for the frequency-dependent penetration depth
$\lambda(T,\Omega)$ may be obtained in the pure regime, $T \gtwid T^*$,
by neglecting self-consistency in the imaginary part of the conductivity
as well,
$$
\Big({{\lambda(T,0)}\over{\lambda(T,\Omega)}}\Big)^2
\simeq\Biggl[1+\Bigl({\lambda(T,0)\over\lambda_0}\Bigr)^2\int 
d\omega\, N(\omega)\, \left(-{\partial
f\over\partial\omega}\right)\Bigl[{(\Omega\tau)^2\over
1+(\Omega\tau)^2}\Bigr]\Biggr].\eqno()
$$ 
In the collisionless limit $\Omega\tau\gg 1$, the response of the
system is perfectly diamagnetic in this approximation,
$\lambda(T,\Omega)\rightarrow \lambda_0$.  In Fig.~6, we explicitly
illustrate the effect of increasing the microwave frequency on the skin
depth of a clean $d_{x^2-y^2}$ superconductor.  \noindent
%\vskip .5cm
\sidehead{\bf IV. Spin fluctuation model for quasiparticle relaxation}
%\vskip 1cm
As discussed in Sec.~III, in the ``pure'' limit where $T^*\ll T \ll
T_c$, we find a ``Drude''-like form (19) for the conductivity of a
d-wave superconductor, with $\tau^{-1}(\omega) =
-2\,\im\,\Sigma_0(\omega)$ and $N(\omega)$ the superconducting density
of states.  In this limit the penetration depth for a $d_{x^2-y^2}$
state is given by
$$
\left({\lambda(0)\over\lambda(T)}\right)^2 = 1 - \int^\infty_{-\infty}
d\omega\, N(\omega)\, \left(-{\partial
f\over\partial\omega}\right). 
\eqno()
$$ 
Then using $(\lambda(0)/\lambda(T))^2 = 1-n_{qp}(T)/n$ to define a
normal quasiparticle fluid density, $\sigma$ may be written as
$$
\sigma_{xx}(\Omega) = {n_{qp}(T)e^2\over
m}\,\im\left\langle{1\over\Omega-i/\tau(\omega)}
\right\rangle, 
\eqno()
$$
where the average $\left\langle ...\right\rangle$ is defined by
$$
\left\langle A(\omega) \right\rangle = 
{\int d\omega\,N(\omega) \left(-{\partial f\over\partial\omega}
\right)
A(\omega) \over \int d\omega\, N(\omega)
\left(-{\partial f\over\partial\omega}\right)}. 
\eqno()
$$ 
In the limit where $\Omega\tau(\omega)\ll 1$, Eq.~(28) reduces to
$\sigma_{xx} = n_{qp}(T)e^2\langle\tau\rangle/m$.
\vskip .2cm
For a $\dxy$ gap, $n_{qp}(T)$ varies linearly with temperature at low
temperatures.  Thus if the average lifetime $\langle\tau\rangle$ were
constant, $\sigma_{xx}$ would vary linearly with $T$ at low
temperatures.  However, the impurity scattering lifetime is
frequency--dependent due to the frequency dependence of the
single-particle density of states.  In Fig.~7 we show plots of
$\tau^{-1}(\omega)$ versus $\omega$ for the case of a $\dxy$ gap and
various values of the scattering phase shift.  In the unitarity limit we
have
$$
{1\over\tau(\omega)} \simeq 
\Biggl\{ { {2T^*~~~~~~~~~~~~~~~~~~~~~~~~\omega<T^*}
\atop {\displaystyle{\pi^2\Gamma\Delta_0\over
2\omega\ln^2(4\Delta_0/\omega)} ~~~~~~~~~~~~~\omega>T^*} }. \eqno()
$$ 
Thus in the "gapless" regime, $\omega<T^*$, the impurity scattering
rate saturates at $2T^*$ and in the "pure" regime, $\omega >T^*$, $\tau$
varies linearly with $\omega$ to within logarithmic factors.  In this
limit, as discussed in Sec. II, the conductivity rises with increasing
temperature as $T^2$ times logarithmic corrections.  This type of
behavior is characteristic of a $\dxy$ gap and resonant impurity
scattering.  One power of $T$ comes from $n_{qp}(T)$ and the other from
$\langle\tau\rangle$; both ultimately reflect the linear $\omega$
variation of the single-particle energy density of states.

At higher temperatures, inelastic scattering and recombination processes
determine the quasiparticle lifetime.  In models in which the $\dxy$
pairing arises from the exchange of antiferromagnetic
spin-fluctuations,\refto{afnormalstate} it is natural to expect that
antiferromagnetic spin fluctuations rather than phonons provide the
dominant inelastic relaxation mechanism.  Calculations of the
quasiparticle lifetime\refto{Quinlanetal} have been carried out for a
two-dimensional Hubbard model in which the spin-fluctuation interaction
is taken into account by introducing an effective interaction
$$ V(\vec
q, \omega) = { {{3\over 2}\,{\overline U}}\over { 1- {\overline
U}{\chi_0^{BCS}} (\vec q, \omega)} }.\eqno()
$$
Here $\overline U$ is a renormalized coupling, and 
$$
\eqalign{
{\chi_0^{BCS}} (\vec q, \omega)= 
{1\over N} {\sum_p} \Big\{ 
{1\over 2}\bigg[1+
{{\epsilon_{p+q}\epsilon_p+\Delta_{p+q}\Delta_p}\over{E_{p+q}E_p}}
\bigg] {{f(E_{p+q})-f(E_{p})}\over{\omega - (E_{p+q}-E_p)+i0^+}}\cr
+{1\over 4}\bigg[1-
{{\epsilon_{p+q}\epsilon_p+\Delta_{p+q}\Delta_p}\over{E_{p+q}E_p}}
\bigg] {{1-f(E_{p+q})-f(E_{p})}\over{\omega + (E_{p+q}+E_p)+i0^+}}\cr
+{1\over 4}\bigg[1-
{{\epsilon_{p+q}\epsilon_p+\Delta_{p+q}\Delta_p}\over{E_{p+q}E_p}}
\bigg] {{f(E_{p+q})+ f(E_{p})-1}\over{\omega - (E_{p+q}+E_p)+i0^+}}
\Big\}
}\eqno()
$$ 
is the BCS susceptibility with
%$\Delta_p = a \Delta_0(T) 
%(\cos p_x - \cos p_y)$
%the $d_{x^2-y^2}$ gap.  In Eq.~(~~~), 
$E_p=\sqrt{{\epsilon_p}^2+{\Delta_p}^2}$, where $\epsilon_p=-2t (\cos
p_x + \cos p_y)-\mu$ With the interaction given by Eq.~(31), the
lifetime of a quasiparticle of energy $\omega$ and momentum $\vec p$ in
a superconductor at temperature $T$ is given to leading order by
$$
\eqalign{
\tau^{-1}_{in}({\vec p}, \omega) 
={1\over N}
{\sum_{p^\prime}} ~~~~~~~~~~~~~~~~~~~~~~~~~~~~~~~~~~~~~~~~~~~~
~~~~~~%
\cr\Big\{\int_0^{\omega-|\Delta_p^\prime|}
d\nu ~~{\rm Im} ~V(p-p^\prime,\nu)\delta(\omega-\nu-
E_{p^\prime})
\Big ( 1 + {{\Delta_p\Delta_{p^\prime}}\over{\omega(\omega-\nu)}} 
\Big ) (n(\nu) +1) [1-f(\omega-\nu)]+\cr
+\int_{\omega+|\Delta_p^\prime|}^0
d\nu ~~{\rm Im} ~V(p-p^\prime,\nu)\delta(\nu-\omega-
E_{p^\prime})
\Big ( 1 -{{\Delta_p\Delta_{p^\prime}}\over{\omega(\nu-\omega)}} 
\Big ) (n(\nu) +1) f(\nu-\omega)+\cr
+\int_0^{\infty}
d\nu ~~{\rm Im} ~V(p-p^\prime,\nu)\delta(\omega+\nu-
E_{p^\prime})
\Big ( 1 + {{\Delta_p\Delta_{p^\prime}}\over{\omega(\omega+\nu)}} 
\Big ) n(\nu) [1-f(\omega+\nu)]\Big \}
}\eqno()
$$ 
Here $n(\nu )$ and $f(\omega )$ are the usual Bose and Fermi factors,
and a quasiparticle renormalization factor has been absorbed into $V$.
The second term of Eq.~(32) corresponds to a process in which two
quasiparticles recombine to form a pair with excess energy emitted as a
spin fluctuation.  The first and third terms describe scattering
processes associated with the emission or absorption of spin
fluctuations, respectively.
\vskip .2cm
Quinlan \etal \refto{Quinlanetal} numerically evaluated Eq.~(31) to
obtain the quasiparticle lifetime using parameters for ${\overline
U},t$, and the band filling which had previously provided a basis for
fitting the nuclear relaxation rate of YBCO\refto{tau_sf} and gave a
normal state quasiparticle lifetime $\tau^{-1}(T_c)$ of order $T_c$.
The temperature dependence of the inelastic quasiparticle lifetime for a
$d_{x^2-y^2}$ gap with $2\Delta_0/T_c = 6$ to $8$ was found to be in
reasonable agreement with the higher temperature transport lifetime
determined by Bonn \etal.  At reduced temperatures below $T/T_c$ of
order 0.8, the $d_{x^2-y^2}$ gap is well-established and the occupied
quasiparticle states are near the nodes.  Setting $\vec p$ to its nodal
value and $\omega=T$, Quinlan \etal\ found that the temperature
dependence of the numerical calculations of the quasiparticle lifetime
varied as $T^3$, reflecting the available phase space.

Figure 8 incorporates results for $\langle\tau\rangle$ obtained by
setting the scattering rate equal to the sum of the impurity and
inelastic rates.  This procedure neglects the real parts of the
self-energy as well as vertex corrections arising from the dynamic
processes.  Nevertheless, it shows the qualitative behavior of
$\langle\tau\rangle$ versus $T/T_c$.  Combining a simple parameterized
fit of the numerical results of
\ref{Quinlanetal} for
$\tau^{-1}_{in}(T)$ with the unitary elastic scattering rate,
corresponding results for $\sigma(T)$ versus $T/T_c$ are shown in
Fig.~9.  Here the peak in $\sigma(T)$ arises from the rapid drop in the
dynamic quasiparticle scattering rate as the gap opens below $T_c$ and
spectral weight is removed from the spin-fluctuations.\refto{Nussetal}
The low-temperature $T^2$ dependence implies that at these energies, the
quasiparticle scattering rate is increasing as the temperature is
lowered due to the linear decrease in the single-particle density of
states and the fact that $\tau$ is proportional to this density of
states in the unitary scattering limit.\refto{HPSPRL} As the microwave
frequency $\Omega$ is increased, the temperature $T_p$, at which the
peak in $\sigma(\Omega,T)$ occurs, increases.  At the same time the peak
value decreases.  Adding the numerical results for the inelastic
scattering rate $\tau^{-1}_{in}(T)$ to the unitary elastic scattering
rate and evaluating Eq.~(25) for various microwave frequencies, we find
that $T_p/T_c$ and $\sigma(\Omega,T_p)/\sigma(0,T_c)$ vary with $\Omega$
as shown in Fig.~10.
%\vskip 1.cm
\sidehead{\bf V. Analysis}
%\vskip .5cm
     Quantitative comparison of the simple theory presented here with
existing data is useful but dangerous.  We remind the reader that many
features of the model are certainly oversimplified, including but not
limited to the neglect of the real Fermi surface anisotropy,
higher-order impurity scattering channels, and strong coupling
corrections.  However, we do not expect inclusion of these aspects of
the physics to qualitatively alter the nature of the temperature power
laws in the response functions at low temperatures in the gapless and
pure regimes.  At higher temperatures $T\ltwid T_c$, it is natural to
expect that real-metals effects will produce nonuniversal behavior in
the superconducting state even if the normal state is a strongly
renormalized Fermi liquid. With these remarks in mind, we proceed as
follows.  We first attempt to fix the impurity scattering parameters
within the resonant scattering model by comparison to the penetration
depth data of Bonn et al.\refto{Bonnhardybigpaper} on Zn-doped samples
of YBCO.  It turns out the fit obtained is relatively good in this case,
although the scattering rates in the case of the Zn-doped samples are
not fixed with high accuracy because of uncertainties in the zero-T
penetration depth.  As discussed below, a different kind of scaling
analysis can be performed on the thin film data of Lee
\etal\refto{LeeLemberger}

\vskip .2cm
As one knows from the heavy fermion superconductivity problem, claims to
determine the gap symmetry by fitting a theoretical prediction to a
single experiment on a single sample should be treated with caution.  It
is extremely important to correlate results on different kinds of
measurements on different samples.  The results of the British Columbia
group afford an excellent opportunity to do this kind of cross-checking.
We therefore adopt for the moment the ``best'' results for the
scattering parameters in the pure and Zn-doped samples from the
penetration depth analysis, and use them to compare calculated
conductivities and surface resistances with the Bonn \etal\
data.\refto{Bonnhardybigpaper} The behavior of the temperature-dependent
conductivity is much richer than that of the London penetration depth,
so it will be important for the consistency of the theory to see which
aspects can be reproduced by the d-wave plus resonant scattering (plus
inelastic scattering) model.
\vskip .2cm
In Fig.~11, we show one possible fit to the UBC penetration depth
data.\refto{Bonnhardybigpaper} The curves represent the theoretical
penetration depth $\lambda(T)$ normalized to the pure London depth
$\lambda_0$ for different values of the resonant scattering parameters
$\Gamma$ as given.  The value $\Delta_0/T_c=3$ is chosen from the fit of
the asymptotic pure $d_{x^2-y^2}$ penetration depth
$\Delta\lambda(T)\simeq\lambda_0 \ln 2 (T/\Delta_0)$ to the intermediate
linear-$T$ regime in the pure data (symbols).  The value
$\Gamma/T_c=8\times 10^{-4}$ is then chosen by fitting the curvature of
the $T^2$ contribution at the lowest temperatures.  As the absolute
scale of the experimental $\lambda(T=0)$ is uncertain, we have chosen to
add constant offsets to the various data sets to try to achieve
reasonable fits.  Figure 11 shows that it is possible to find a
consistent choice of such offsets, since the scattering rates used for
the two Zn-doped data sets, $\Gamma/T_c=0.018$ and $0.009$ are in the
ratio 2:1 as are the nominal Zn concentrations 0.31\% and 0.15\%.
However, a roughly equally good fit may be obtained using scattering
rates of, e.g., $\Gamma/T_c=0.03$ and $0.006$, which would then not be
consistent with the theoretically predicted scaling of $\Gamma$ with the
impurity concentration $n_i$.
Clearly there is a relatively large range of acceptable scattering rates
corresponding to the two Zn-doped curves, possibly a factor of two or
more.  A determination of the zero-temperature limiting penetration
depths of pure and Zn- doped samples from, \eg\ $\mu SR$ experiments, is
needed to fix these values more precisely or rule out such a fit.
\vskip .2cm
A procedure for fixing the zero-temperature penetration depth relative
to the single-crystal data without new experiments has been suggested by
Lee \etal\ They assume that the data for their YBCO films follow a
universal curve given by the form of the single crystal penetration
depth in the intermediate temperature regime, as suggested by the
resonant scattering analysis.  Using data on several films, they show
that such a scaling is indeed possible, and assign zero-temperature
penetration depth values to several films on this basis.  This allows an
internal consistency check of the resonant scattering hypothesis,
wherein one may check to see that the measured coefficients of the $T^2$
term in the penetration depth, equal to $c_2=\pi\lambda_0/(6\gamma
\Delta_0)$ for a $d_{x^2-y^2}$ state and resonant scattering, scale
appropriately with the zero-temperature penetration depth
renormalization,
\break
$({\tilde\lambda}_0-\lambda_0)/\lambda_0\simeq (\gamma/(\pi\Delta_0))
\ln(4\Delta_0/\gamma)\simeq\Gamma /(2\gamma)$.  Since a given film
in the resonant scattering limit is characterized simply by its impurity
concentration through the parameter $\gamma$, using the above
expressions it is possible to check scaling without knowledge of the
actual defect concentration.  For example, in Fig.~12 we plot
$({\tilde\lambda}_0-\lambda_0)$ vs. $1/c_2$ for two ``different'' films
measured in \ref{LeeLemberger} actually the same film before and after
annealing (films A and A$'$ of
\ref{LeeLemberger} ).
Each cluster of points in Figure 12 represents a single film, the
individual points corresponding to differing assumptions regarding other
constants, such as the absolute value of the pure penetration depth,
which enter such an analysis.  It is seen that the agreement with the
theoretical scaling is remarkably good, and that this agreement is not
particularly sensitive to varying assumptions on the subsidiary
constants.
\vskip .2cm

Next we explore whether an equally good fit is possible for the
resistive part of the conductivity which was also measured in
\ref{Bonnhardybigpaper}.  As we have seen, even in the "pure" limit 
$T>T^*$ the conductivity depends on the quasiparticle lifetime.  At low
temperatures, elastic scattering from impurities determines this
lifetime.  At higher temperatures, however, inelastic scattering
processes become important and we use a simple parameterized fit to the
numerical results for the inelastic scattering rate $\tau^{- 1}(T)$
obtained by Quinlan et al.\refto{Quinlanetal} As previously discussed,
the parameters of the spin-fluctuation interaction used in this work
were used in fitting the NMR data and the overall strength was adjusted
to give $\tau_{in}^{- 1}(T_c)$ of order $T_c$.  The total scattering
rate is taken as the sum of the elastic and inelastic rates.  Using the
usual expression, for the surface resistance $R_s$ in terms of the real
part of the conductivity $\sigma$ and the penetration depth,
$R_s=(8\pi^2\Omega^2\lambda^3\sigma)/c^4$, Bonn \etal\ extracted the
conductivity for the same samples whose penetration depth is plotted in
Fig.~11.  In Figs.~13 and 14, we show the conductivity plotted for these
samples calculated using the elastic scattering parameters taken from
Fig.~10 and the inelastic scattering results from Fig. 8.  Although the
size, position, and scaling with frequency of the prominent maximum in
the conductivity are reproduced qualitatively, it is clear that the
low-temperature behavior of the data does not correspond to the
predictions of the model.  In section II, we pointed out that, while a
$\sigma\sim T$ behavior can be obtained in the pure regime if
$\Omega\tau\simeq 1$, it is not generic to the theory; by contrast, the
data for at least the "pure" sample and 0.15\% Zn appear to follow a
low-temperature linear-$T$ law for all the samples shown.  A similar
behavior is observed in YBCO thin films and BSSCO single
crystals.\refto{Maetal}
\vskip .2cm
The further difficulty apparent from the data shown in Figs.~13 and 14
is the rather large residual value of the conductivity as $T\rightarrow
0$ exhibited by all data sets.  While the $d$-wave theory predicts a
residual absorption, the limiting $\sigma_{\scriptscriptstyle00}\simeq
ne^2/m\pi\Delta_0$ of the theory is an order of magnitude or so lower
than that extracted by the British Columbia
group.\refto{Hardy3,Bonnhardybigpaper} While qualitatively different
physical scattering mechanisms than those considered here, or a
completely different picture for superconductivity in the cuprates might
be responsible for the deviations from theory apparent in the data, we
prefer to reserve judgement until further data is available.  Very
recent results from the British Columbia group indicate that twin
boundaries may be responsible for the residual conductivities observed,
and possibly also account for part of the temperature dependence
observed at low temperatures.  In Fig.~15 we show data for a twin-free,
high-purity YBCO crystal\refto{Bonnhardybigpaper} compared to the same
theoretical prediction used for the low-frequency conductivity displayed
in Fig.~13.  It is evident that the residual conductivity in the
untwinned has been dramatically reduced, and the low-temperature fit to
the $d$-wave theory correspondingly improved.  Clearly high-quality
Zn-doped samples of this type are also desirable.
\vskip .2cm
For completeness we also calculate and display the surface resistance
$R_s(T)$ for various values of the scattering parameters in Figure 16.
Here again, we see that the low-temperature behavior of the theory is in
disagreement with the
data.\refto{Hardy1,Hardy2,Hardy3,Bonnhardybigpaper} This reflects the
much lower residual conductivity predicted for our model, as well as the
$T^2$ power law dependence.  In addition, in order to reproduce the
dramatic decrease in $R_s$ which is observed below $T_c$, we need a
large $\Delta_0/T_c = 4$ ratio.  It is also important in making this
comparison to recall that the drop in $R_s$ just below $T_c$ reflects
less the collapse of the inelastic scattering rate which enters the
conductivity $\sigma$ than the divergence of the penetration depth depth
near $T_c$ (recall $R_s\sim\lambda^3$).  The data suggests that the
magnitude of the gap opens more rapidly than usual.  This type of
behavior has been found in model calculations based on the exchange of
spin fluctuations including processes not considered
here.\refto{Monthoux1,Bickers} It is also possible that critical effects
in a range of up to several degrees near the transition may lead to a
divergence more rapid than in the usual mean field case.\refto{lamfluc}

\sidehead{\bf VI.  Conclusions}
%\vskip 1cm

In this paper we have calculated $\lambda (\Omega,T)$ and
$\sigma(\Omega,T)$ within the framework of a BCS model in which the gap
has $\dxy$ symmetry, and both strong elastic impurity scattering and
spin-fluctuation inelastic scattering processes are taken into account.
We have sought to address a set of basic questions raised in the
introduction.  Here we summarize what we have learned.

\item{1)} The microwave conductivity of the layered cuprates can be
written in a Drude-like form
$$
\sigma(\Omega,T) = {n_{qp}(T)e^2 \over m}\>\im\,\left\langle
{1\over\Omega-(i/\tau(\omega,T))}\right\rangle.\eqno()
$$ 
Here $n_{qp}(T)$ is the normal quasiparticle fluid density and the
brackets denote the frequency average defined in Eq. (28).  The inverse
quasiparticle lifetime $\tau^{-1}(\omega,T)$ is the sum of the elastic
impurity scattering rate and the inelastic spin-fluctuation scattering.
The form of Eq. (28) describes the transport properties of nodal
quasiparticles which have a relaxation time $\tau(\omega,T)$ and a
density of states $N(\omega)$.

\item{2)} In the hydrodynamic limit  $\Omega\langle\tau\rangle \ll 1$,
$\sigma(T) = $ $n_{qp}(T)e^2\langle\tau\rangle/ m.$ This is just the
form that Bonn \etal\ used to extract a quasiparticle lifetime from
their conductivity data.  Here we have shown that $\langle\tau\rangle$
corresponds to an average over a frequency- and temperature-dependent
lifetime.  Figure 8 shows a plot of $\langle\tau\rangle^{-1}$ versus $T$
for typical parameters.

\def\soo{\sigma_{\scriptscriptstyle 00}}
\item{3)} We find that for a $\dxy$ gap, $\sigma(T\to0)$ goes to a
constant $\soo=ne^2/m\pi\Delta_0$ independent of the impurity
concentration (for small concentrations).\refto{PALee} If we take
$\tau^{-1}(T_c)\simeq T_c$ from DC resistivity measurements, and
$2\Delta_0/kT_c=6$, then $\soo/\sigma(T_c)=1/3\pi$ so that the limiting
value of $\soo$ is a about an order of magnitude smaller than
$\sigma(T_c)$.  As the temperature increases, $\sigma(T)$ grows as
$T^2$.  For $T>T^*$, this can be understood as arising from the fact
that both $n_{qp}(T)$ and $\langle\tau\rangle$ in the resonant
scattering limit vary linearly with $T$.  Note that we also find that
for $T<T^*$, $\sigma(T)-\soo$ varies as $T^2$.  If, in the pure limit
$T>T^*$, $\langle\tau\rangle$ were a constant, then $\sigma(T)$ would
increase linearly with $T$.  However, this is not the case for the model
we have considered.  Both the fact that $\sigma(T\to0)$ is independent
of the impurity concentration and that $\sigma(T)$ increases as $T^2$
appear to be in disagreement with the presently available data.  There
is some evidence that the residual conductivity may be substantially
lowered by reducing the density of twins in the
crystal,\refto{Bonnhardybigpaper} but the linear-T behavior remains a
puzzle. Whether other scattering mechanisms can give rise to this
behavior is not at present understood.  The effect of particle-hole
asymmetry is of particular interest in the context of our observation
that a constant relaxation time at low temperatures in pure samples is
needed to produce a linear temperature dependence.  The analytic
properties of the self-energy of a particle-hole symmetric
superconductor formally preclude such a result, however.  An
investigation of particle-hole asymmetry effects is in progress.

\item{4)} At higher temperatures, inelastic scattering processes 
become important and give rise to a scattering rate which increases
initially as $(T/T_c)^3$.  As shown in Fig.~8, this leads to a minimum
in $\langle\tau\rangle^{-1}$ at a particular value of $T/T_c$.

\item{5)} At higher microwave frequencies where
$\Omega\langle\tau\rangle\sim1$, there is a crossover from the
hydrodynamic to the collisionless regime, and the relationship of
$\sigma(T,\Omega)$ to the quasiparticle lifetime involves an average of
$\tau(\omega,T)/(1+\Omega^2\tau^2(\omega,T))$.  In this regime, the
conductivity can exhibit a quasi-linear variation with $T$.  We have
shown in Fig.~10 how the temperature $T_p$ of the peak conductivity
varies with $\Omega$ along with $\sigma(\Omega,T_p)/\sigma(T_c)$.  We
have also found that at higher microwave frequencies, quasiparticle
screening leads to a reduction in $\lambda(T,\Omega)$.  At a fixed
temperature $\lambda(T,\Omega)$ can approach $\lambda(0,0)$ as $\Omega$
increases.  We have used the full frequency dependence of
$\lambda(T,\Omega)$ and $\sigma_1(T,\Omega)$ in calculating the surface
resistance $R_s(T,\Omega)$ shown in Fig.~16.

\item{6)} In Section V, we explored the extent to which the $\dxy$-
wave plus scattering model can describe the surface impedance observed
in YBa$_2$Cu$_3$O$_{6.95}$ and its Zn-doped variants.  It appears
(Figs.~11 and 12) that the temperature- and impurity- dependence of the
penetration depth can be fit within the framework of this model.  It
will be interesting to compare the results for the $\Omega$-dependence
of $\lambda(T,\Omega)$ with experimental results which will soon be
available.\refto{Kleinorensteinprivate} The measured values of
$\sigma_1(T,\Omega)$ shown in Fig.~14 for the pure and 0.15$\,$\% Zn
samples appear to have a linear low-temperature variation in contrast to
the $T^2$ variation predicted from the model.  In addition, as noted,
the limiting residual value of the conductivity obtained from the theory
is smaller than that observed in many samples and is independent of the
concentration of impurities.  Nevertheless, as shown in Figs.~11 and 15,
a simple $d$-wave model plus scattering provides a reasonable overall
fit to both the real and imaginary parts of the conductivity.  One can
ask whether alternative models such as an anisotropic $s$-wave pairing
could provide similar fits to the data.  In the absence of impurity
scattering, the penetration depth and the low-frequency microwave
conductivity $\sigma(T)$ will both vary exponentially at temperatures
below the minimum gap value.  In addition, if the minimum gap value is
finite, $\sigma(T\to0)$ will vanish as $\exp -(\Delta_{min}/T)$.  An
extreme example of an anisotropic $s$-wave gap is given by taking for
$\Delta$ the magnitude of the $\dxy$ gap, $\Delta_0(T) |\cos 2\phi|$.
In this case, the results in the pure limit for $\lambda(T)$ are
identical to the $\dxy$ results.  However, the addition of impurities
can lead to a qualitatively different behavior for the anisotropic
$s$-wave case.\refto{LBSPJH} As discussed in Section II, both
$\tilde\omega_n$ and $\tilde\Delta_k$ are renormalized by impurities in
the $s$-wave case.  In particular, potential scattering acts to average
the gap over the Fermi surface, thus reducing the peak value of the gap
and increasing the minimum value.  Thus, even if one took the extreme
anisotropic $s$-wave case in which the gap has nodes but does not change
sign, impurities would lead to a finite effective gap and an exponential
rather than $T^2$ crossover of the low-temperature dependence of both
$\lambda (T)$ and $\sigma(T)$.  If "inert" defects like Zn impurities
are found to have a magnetic character,\refto{Alloul} however,
distinguishing $s-$ and $d-$wave states becomes more
difficult.\refto{LBSPJH} Further measurements of the low-temperature
dependence of the surface impedance in pure and impurity doped cuprates
along with detailed comparisons with theoretical models are necessary to
determine the symmetry of the pairing state.

\medskip
\centerline{\bf Acknowledgments}
\smallskip
The authors acknowledge extremely useful discussions with D.A.~Bonn,
W.N.~Hardy, and T.~Lemberger, and are grateful to them for sharing their
data prior to publication.  S. Quinlan generously provided assistance
with calculations of the inelastic quasiparticle lifetime.  PJH is
grateful to N. Goldenfeld for many enlightening exchanges.  DJS was
supported in part by the National Science Foundation under grant
DMR92--25027.  WOP was supported by NSF DMR91--14553 and by the National
High Magnetic Field Laboratory at Florida State University.  Numerical
Computations were performed on the Cray-YMP at the Florida State
University Computing Center.
\vskip 1cm

\references
\refis{Hardypendepth} W.N. Hardy, D.A. Bonn,  D. C. Morgan, R. Liang, 
and K. Zhang, Phys. Rev. Lett. 70, 399, (1993).

\refis{LeeLemberger}J.Y. Lee, K. Paget, T. Lemberger, S. Foltyn 
and X. Wu, preprint  (1993).

\refis{Grossetal}F. Gross , B.S. Chandrasekhar, D. Einzel, K. Andres,
P.J. Hirschfeld, H.R. Ott, J. Beuers, Z. Fisk and J.L. Smith, Z. Phys.
64, 175 (1986).

\refis{NMR}K. Ishida Y. Kitaoka, T. Yoshitomi, N. Ogata, T. Kamino, 
and K. Asayama, Physica C179, 29 (1991); J.A. 
Martindale, S.E. Barrett, C.A. Klug, K.E. O'Hara, S.M. DeSoto, 
S.P. Slichter, T.A. Friedmann and D.M. Ginsberg, Phys. Rev. Lett. 68, 
702 (1992); K. 
Ishida, Y. Kitaoka, N. Ogata, T. Kamino, K. Asayama, J.R. Cooper, 
and N. Athanassopoulou,  J. Phys. Soc. Japan
62, 2803 (1993).

\refis{photo}Z. Shen,  D.S. Dessau, B.O. Wells, D.M. King, W.E. Spicer, 
A.J. Arko, D. Marshall, L.W. Lombardo, A. Kapitulnik, P. Dickinson, 
S. Doniach, J. DiCarlo, T. Loeser, and C.H. Parks, Phys. Rev. Lett. 
70, 1553 (1993); R.J. Kelley, J. Ma, G. Margaritondo, and M. Onellion,  
preprint (1993).

\refis{squid} D.A. Wollman, D.J. van Harlingen, W.C. Lee, D.M. Ginsberg,
and A.J. Leggett, Phys. Rev. Lett. 71, 2134 (1993).

\refis{Carbotte} M. Prohammer and J. Carbotte, Phys. Rev. B43, 
5370 (1991); P. Arberg, M. Mansor and J.P. Carbotte, Sol. St. Commun.
86, 671 (1993).

\refis{felds} P.J. Hirschfeld and N. Goldenfeld, Phys. Rev. B48 (1993).

\refis{HPSPRL} P.J. Hirschfeld, W.O. Putikka, and D.J. Scalapino,
Phys. Rev. Lett. 71, 3705 (1993).

\refis{DJSq=0} J.-J. Chang and D.J.~Scalapino, \prb 40, 4299, 1989.

\refis{KimMuzikar} H. Kim, G. Preosti, and P. Muzikar, preprint (1993).

\refis{ChoiMuzikar} C.H. Choi and P. Muzikar, Phys. Rev. B39, 11296
(1989).

\refis{Hardy1} D.A. Bonn, R. Liang, T.M. Riseman, D.J. Baar, 
D.C. Morgan, K.
Zhang, P. Dosanjh, T.L. Duty, A. MacFarlane, G.D. Morris, J. H. Brewer,
W.N. Hardy, C. Kallin, and A.J. Berlinsky, Phys. Rev. B47, 11314 (1993).

\refis{Hardy2} D.A. Bonn, K. Zhang, R. Liang, D.J. Baar, and W.N. Hardy,
J. Supercond. 6, 219 (1993).

\refis{Hardy3} D.A. Bonn, D.C. Morgan, K. Zhang, R. Liang, D.J. Baar, 
and W.N. Hardy, J. Phys. Chem Sol. 54, 1297 (1993); 
K. Zhang, D.A. Bonn, R. Liang, D.J. Baar,  and  W.N. Hardy, 
Appl. Phys. Lett. 62, 3019 (1993).

\refis{Nussetal} M.C. Nuss, P.M. Mankiewich, M.L. O'Malley, 
E.H. Westerwick, and P.B. Littlewood, Phys. Rev. Lett. 66, 3305 (1991).  

\refis{PALee} 
         P.A. Lee, Phys. Rev. Lett. 71, 1887 (1993).

\refis{Klemm1} 
        R.A Klemm, K. Scharnberg, D. Walker, and C.T. Rieck,  
Z. Phys. 72, 139 (1988).% #2

\refis{PJHcond} 
P.J. Hirschfeld, P. W\"olfle, D. Einzel, J.A. Sauls, and W.O. Putikka,
Phys. Rev. 40, 6695 (1989).%

\refis{PJHconsequences} P.J. Hirschfeld, D. Einzel, and P. W\"olfle,
Phys. Rev. B.37, 83 (1988).

\refis{Arfi} B. Arfi, H. Bahlouli, C.J. Pethick, and D. Pines, Phys.
Rev. Lett. 60, 2206 (1988).

\refis{PJHskin}
 P.J. Hirschfeld, W.O. Putikka, P. W\"olfle and Y. Campbell,
 J. Low Temp. Phys. 88, (1992); erratum to be published, ibid (1994).

\refis{Putikka1} See also W.O. Putikka and P.J. Hirschfeld,  
to be published in Proc. XX Int. Conf. on Low Temp. Phys. (Eugene, 1993).

\refis{Coffey} L. Coffey, T.M. Rice, and K. Ueda, J. Phys. 
C18, L813 (1985).

\refis{PethickPines} C.J. Pethick and D. Pines, 
Phys. Rev. Lett. 50, 270 (1986). 

\refis{Quinlanetal} S. Quinlan, D.J. Scalapino, and 
N. Bulut, preprint (1993).

\refis{afnormalstate} N.E. Bickers, D.J. Scalapino, and S.R. White, 
Phys. Rev. Lett. 62, 961 (1989); P. Monthoux and D. Pines, Phys. Rev. 
Lett. 69, 961 (1992).

\refis{tau_sf} N. Bulut and D.J. Scalapino, Phys. Rev. Lett. 68, 706 
(1992); D. Thelen, D. Pines, and J.P. Lu, Phys. Rev. B47, 915 (1993).

\refis{AGD} A.A. Abrikosov,  L.P. Gor'kov, and  I. E. Dzyaloshinski, 
{\it Methods of Quantum Field Theory in Statistical Physics}, (Dover
Books: New York, 1963).

\refis{Skalskietal} S. Skalsi, O. Betbeder-Matibet, and P. Weiss,
Phys. Rev. 136, A1500 (1964).

\refis{dwavereviews} For reviews 
see J. Annett, N. Goldenfeld and S. Renn,  
Phys. Rev. B43, 2778 (1991); D. Pines, Proceedings of the Conference 
on Spectroscopies on Novel Superconductors, Los Alamos, 1993
to appear in J.\ Chem.\ Phys.\ Solids; D.J. Scalapino, ibid.

\refis{MH} See, e.g. E. M\"uller-Hartmann, in {\it Magnetism}, vol. V,
ed. by H. Suhl, (Academic Press: New York, 1973), and references
therein.

\refis{PJHresonant} P.J. Hirschfeld, D. Vollhardt, and P. W\"olfle,
Solid State Commun. {\bf 59}, 111 (1986).

\refis{SchmittRinketal}  S. Schmitt-Rink, K. Miyake, and C.M. Varma,
Phys. Rev. Lett. {\bf 57}, 2575 (1986).

\refis{Kaplanetal}  S.B. Kaplan, C.C. Chi, D. N. Langenberg, J.J. Chang,
S. Jafarey, and D.J. Scalapino, Phys. Rev. B14, 4854 (1976).

\refis{MattisBardeen} D.C. Mattis and J. Bardeen, Phys. Rev. 111,
412 (1958).

\refis{PJHresbogo} W.O. Putikka, P.J. Hirschfeld, and P. W\"olfle,
Phys. Rev. B41, 7285 (1990).

\refis{Poil} D. Poilblanc, W. Hanke and D.J. Scalapino, in 
Phys. Rev. Lett. 72, 884 (1994).

\refis{LBSPJH} L. Borkowski and P.J. Hirschfeld,  preprint (1994).

\refis{Bonnhardybigpaper}  D.A. Bonn, S. Kamal, K. Zhang, R.Liang, 
D.J. Baar, E. Klein, and W.N. Hardy,  preprint (1993).

\refis{Walstedt} R. Walstedt et al., Phys. Rev. B48, 10646(1993).

\refis{Alloul} H. Alloul, P. Mendels,  H. Casalta, J.F. Marucco 
and J. Arabski, Phys. Rev. Lett. 67, 3140 (1991). 

\refis{MonthouxPines}P. Monthoux and D. Pines, preprint (1993).

\refis{AG} A.A. Abrikosov and L.P. Gor'kov, JETP 39 (1960), 1781.

\refis{Balatsky} A.V. Balatsky, A. Rosengren, and B.L. Altshuler, 
preprint (1994).

\refis{Maetal} Z. Ma et al., Phys. Rev. Lett 71, 781 (1993); Z. Ma, 
private communication.

\refis{Tsvelick} A.A. Nersesyan, A.M. Tsvelick, and F. Wenger, 
preprint (1994).

\refis{lamfluc} S. Kamal, D.A. Bonn, N. Goldenfeld,  P.J. Hirschfeld,  
R. Liang, and W.N. Hardy, preprint (1994).

\refis{Monthoux1} P. Monthoux and D.J. Scalapino, Phys. Rev. Lett., 
to be published.

\refis{Bickers} C.-H. Pao and N.E. Bickers, Phys. Rev. Lett., 
to be published.

\refis{Bulut3} N. Bulut, D. Hone, D.J. Scalapino, and E.Y. Loh, 
Phys. Rev. Lett. 62, 2192 (1989).

\refis{Kleinorensteinprivate} N. Klein and J. Orenstein, 
private communications.

\endreferences

\vfill\eject
\doublespace
\centerline{\bf Figure Captions}
%\vskip .2cm
\item{1.} Normalized low-T conductivity, $\sigma/\sigma_{00}$ vs. the
reduced temperature $T/T_c$ for microwave frequency $\Omega=0$.  The
solid lines correspond to resonant scattering, $c=0$, $\Gamma
/T_c=0.01,0.003,0.001$, and dashed line corresponds to $c=0.3, \Gamma
/T_c=0.01$.
%/phys/neptune1/pjh/sigma/fig1.plt
\item{2.} Normalized low-T conductivity, $\sigma/\sigma_{00}$ vs. the
reduced temperature $T/T_c$ in the Born limit, $\Gamma_N/T_c=0.01,
\Omega  /T_c=0,0.001,0.01.  $
%/phys/neptune1/pjh/sigma/fig2.plt
\item{3.} Normalized low-T conductivity, $\sigma/\sigma_{00}$ vs. 
the reduced temperature $T/T_c$ in the resonant limit, for $c=0$,
$\Gamma /T_c=0.001$, and $\Omega /T_c=0,0.0032,0.01. $
%/phys/neptune1/pjh/sigma/fig3.plt
\item{4.} Normalized conductivity, $\sigma/\sigma_{00}$ vs. 
the reduced frequency $\Omega /T_c$ for $T=0$, and $\Gamma
/T_c=0.001,0.01,0.1$.
%no figure exists so far
\item{5.} Normalized zero-temperature London penetration depth, 
$\lambda(T=0)/\lambda_0$ vs. the reduced scattering rate, 
$\Gamma /T_{c0}$ in the resonant scattering limit, c=0.
%/phys/neptune1/pjh/sigma/fig5.plt
\item{6.} Normalized London penetration depth, 
$\lambda(T)/\lambda_0$ vs. the reduced temperature, 
$T/T_c$ for resonant scattering, 
$\Gamma /T_c=0.0008$, $c=0$, and $\Omega  /T_c=0,0.002,0.018$.
%/phys/neptune1/pjh/sigma/fig6.plt
\item{7.} Impurity relaxation rate $1/T_c\tau(\omega)$ vs. 
the reduced frequency $\omega/\Delta_0$ for 
$\Gamma /T_c=0.01,0.001$ and $c=0$
(solid lines) and $\Gamma /T_c=0.01, c=.2$ (dashed line).
%no figure exists yet
\item{8.}  Relaxation rate including inelastic scattering 
$1/T_c\langle \tau\rangle$ vs. the reduced temperature 
$T/T_c$ for $\Gamma /T_c=0.0008,0.009,0.018$, $c=0$, 
$\Delta_0/T_c=3$ (solid lines) and $\Gamma /T_c=0.0008, 
c=0, \Delta_0/T_c=4$ (dashed line).
%no figure exists yet
\item{9.}  Normalized conductivity including inelastic scattering, 
$\sigma/\sigma_{00}$ vs. 
the reduced temperature $T/T_c$ in the resonant limit, $c=0$ for 
$\Omega
/T_c=0.018, c=0,$ and $\Gamma /T_c=0.0008, 0.009,0.018$.
%/phys/neptune1/pjh/sigma/fig9.plt
\item{10.} Reduced conductivity peak temperature,  
$T_p/T_c$  vs.  $\Omega  /T_c$ for $\Gamma /T_c=0.001, 
0.01$, $c=0,$ and $\Delta_0/T_c=3$ (left axis); normalized peak 
conductivity $\sigma(T_p,\Omega  )/\sigma(T_c,0)$ vs. 
$\Omega  /T_c$ (right axis).
%no figure exists so far! 
\item{11.}  Comparison of d-wave penetration depth with penetration 
depth data on YBCO single crystals.\refto{Bonnhardybigpaper}  
Normalized penetration depth,
$\lambda(T)/\lambda_0$ vs. the reduced temperature $T/T_c$ for 
$\Gamma
/T_c=0.018,0.009,0.0008$ and $c=0$.  Data for pure YBCO crystal
(circles), 0.15\% Zn (diamonds), and 0.31\% Zn (squares).
%/phys/neptune1/pjh/dx2-y2/bonnhardy/lambda.dx2-y2.plt
\item{12.}  Normalized T=0  normal fluid density 
$1-(\lambda_0^2/{\tilde\lambda}_0^2)$ vs. the reduced coefficient 
of $T^2$ term, $\lambda_0/(c_2\Delta_0^2)$ in the $\dxy$ plus 
resonant scattering model.  Each cluster of points represents one 
YBCO film from \ref{LeeLemberger}.
%/phys/neptune1/pjh/dx2-y2/lemberger/lembergthy.plt}
\item{13.}  Normalized theoretical conductivity 
$\sigma/\sigma_{1}(T_c)$ vs. the reduced temperature $T/T_c$ for
impurity parameters $\Gamma /T_c=0.0008$ and $c=0$, including inelastic
scattering for $\Omega /T_c=0.002$ and $0.018$ (solid lines).  Data
points are normalized conductivities of YBCO single crystals from
\ref{Bonnhardybigpaper} for microwave frequencies 3.88 GHz (circles) and
34.8 GHz (triangles).
%/phys/neptune1/pjh/dx2-y2/bonnhardy/sigma.rawfit.plt
\item{14.} Normalized theoretical conductivity 
$\sigma/\sigma_{1}(T_c)$ vs. the 
reduced temperature $T/T_c$ for impurity parameters 
$\Gamma /T_c=0.0008, 0.009$ and  $0.018$ with $c=0$, including 
inelastic scattering for $\Omega  /T_c= 0.018$ (solid lines).  
Data points are normalized conductivities of YBCO single crystals from 
\ref{Bonnhardybigpaper} for frequency 34.8 GHz, for samples nominally 
pure (circles), 0.15\% Zn (triangles), and 0.31\% Zn (squares).
%/phys/neptune1/pjh/dx2-y2/bonnhardy/sigmaZn.rawfit.plt
\item{15.}  Effect of detwinning.  Normalized theoretical conductivity 
$\sigma/\sigma_{1}(T_c)$ vs. the 
reduced temperature $T/T_c$ for impurity parameters $\Gamma /T_c=0.0008$
and $c=0$, including inelastic scattering for $\Omega /T_c= 0.002$
(solid line). Data points are normalized conductivities of detwinned
YBCO single crystal from
\ref{Bonnhardybigpaper} for frequency 4.1 GHz. 
%/phys/neptune1/pjh/dx2-y2/bonnhardy/sigma.tw-free.plt
\item{16.} Normalized surface resistance, $R_s/R_s(T_c)$ vs. the reduced
temperature $T/T_c$.  Theory for $\Omega  /T_c=0.002$   and impurity 
parameters $\Gamma /T_c=0.0008, c=0$, including inelastic scattering, 
for $\Delta_0/T_c=3$ (solid line) and $\Delta_0/T_c=4$ (dashed line).  
Data from \ref{Bonnhardybigpaper}, 3.88GHz, nominally pure YBCO crystal. 
\vfill\eject
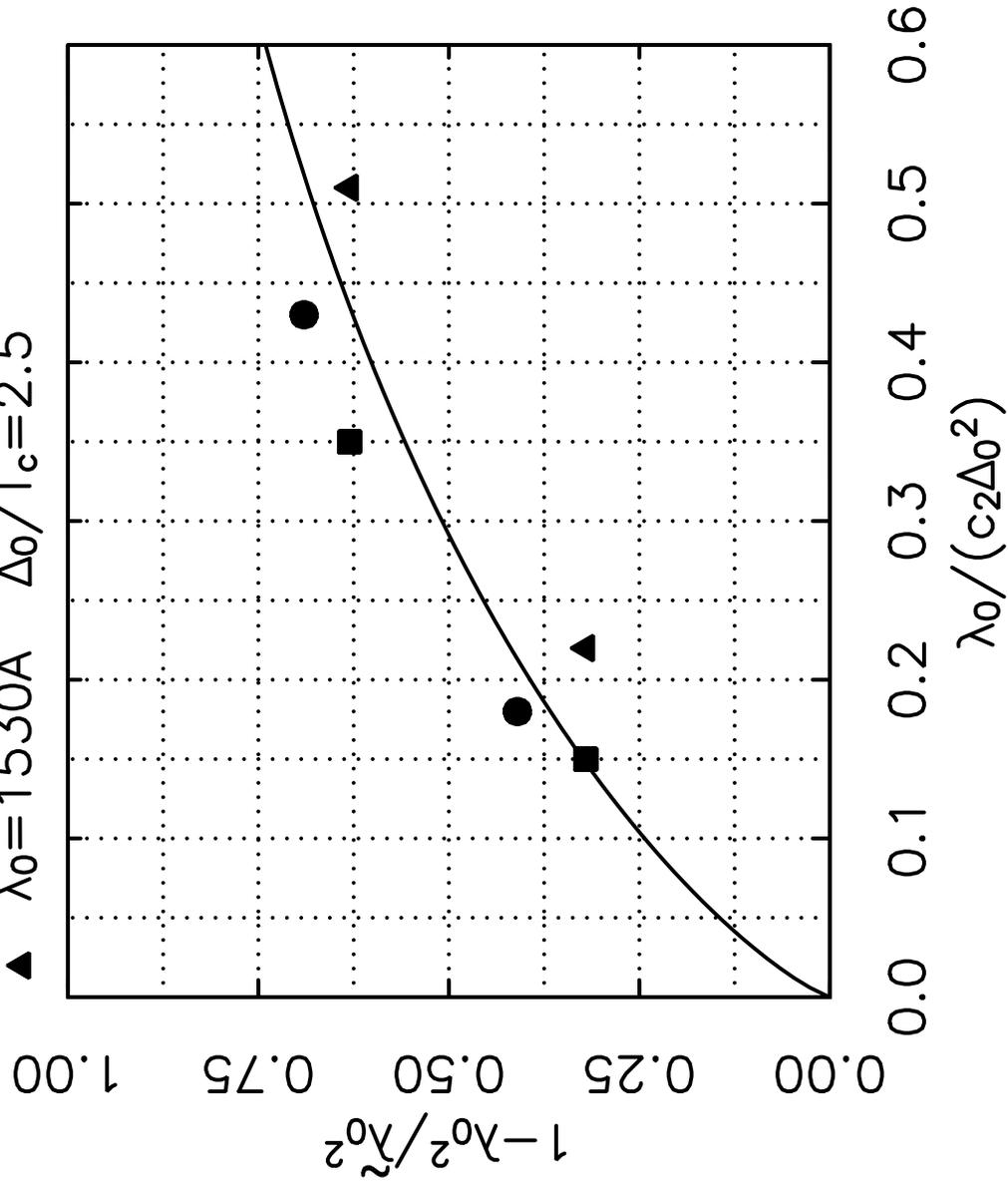
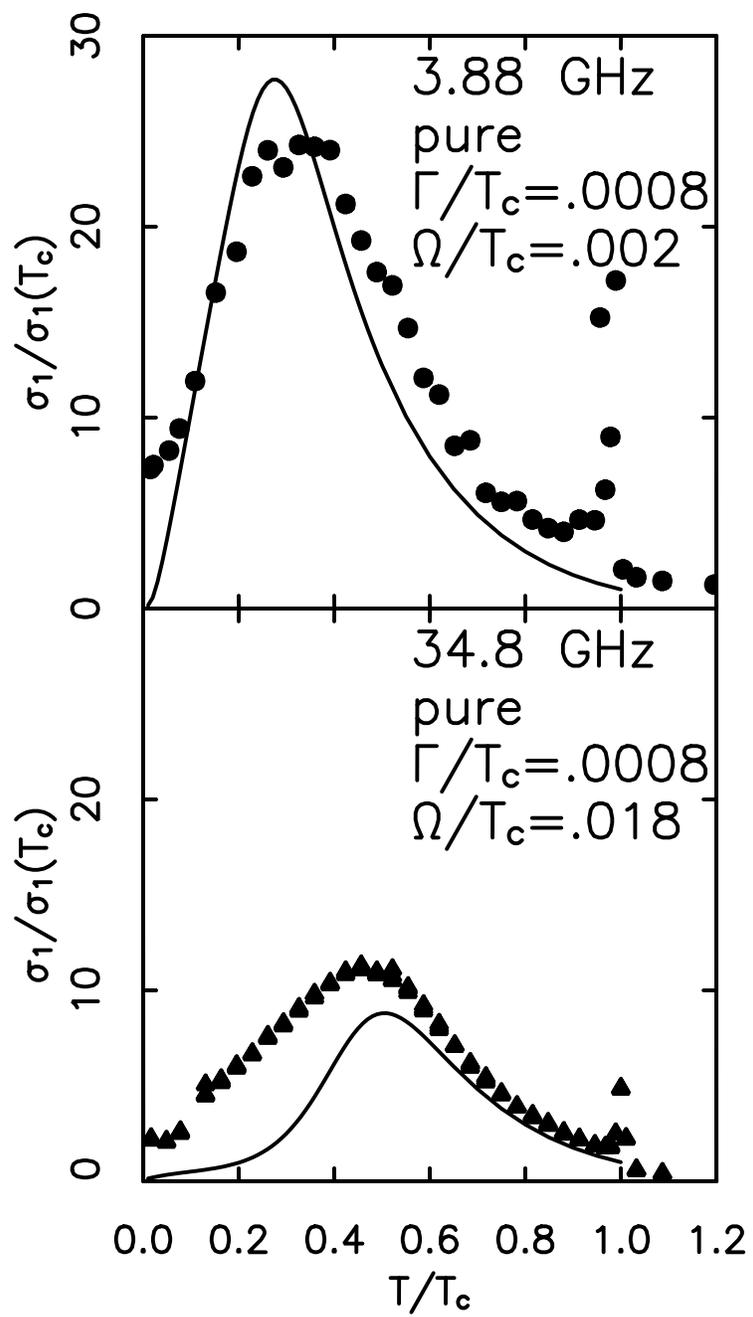
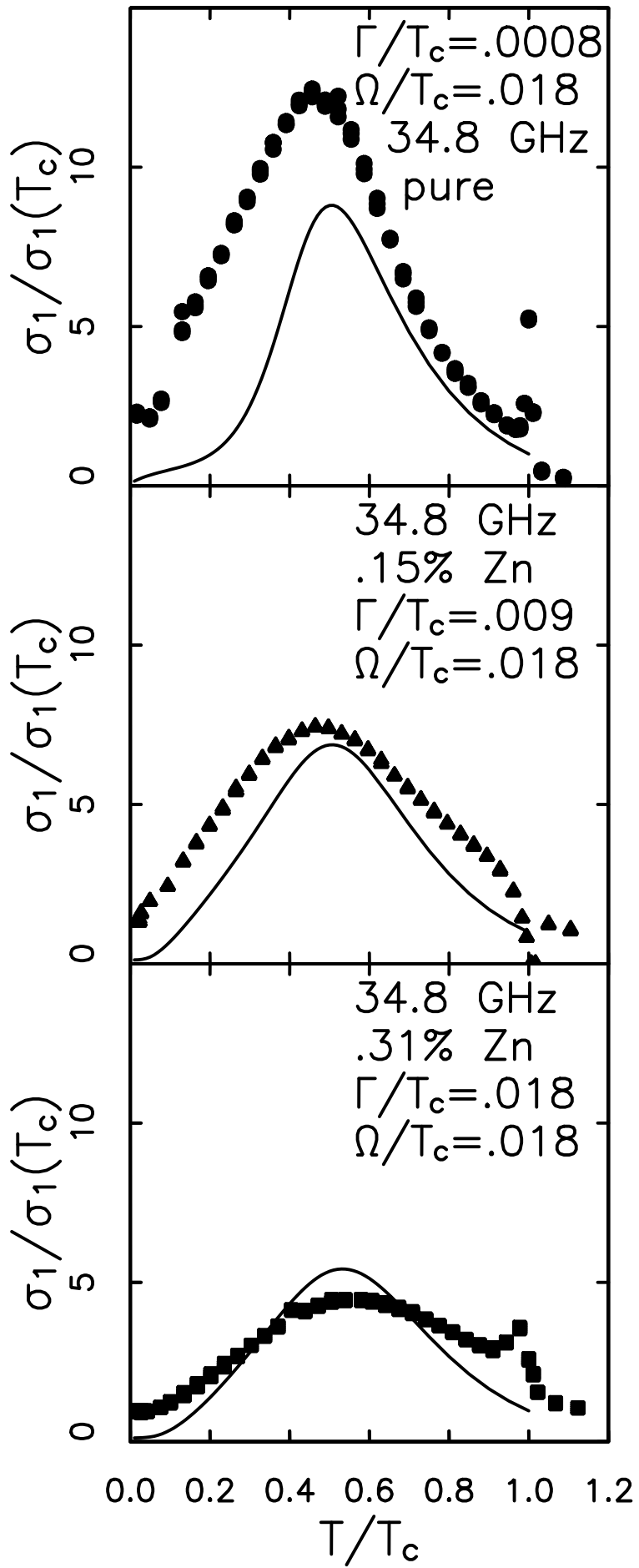
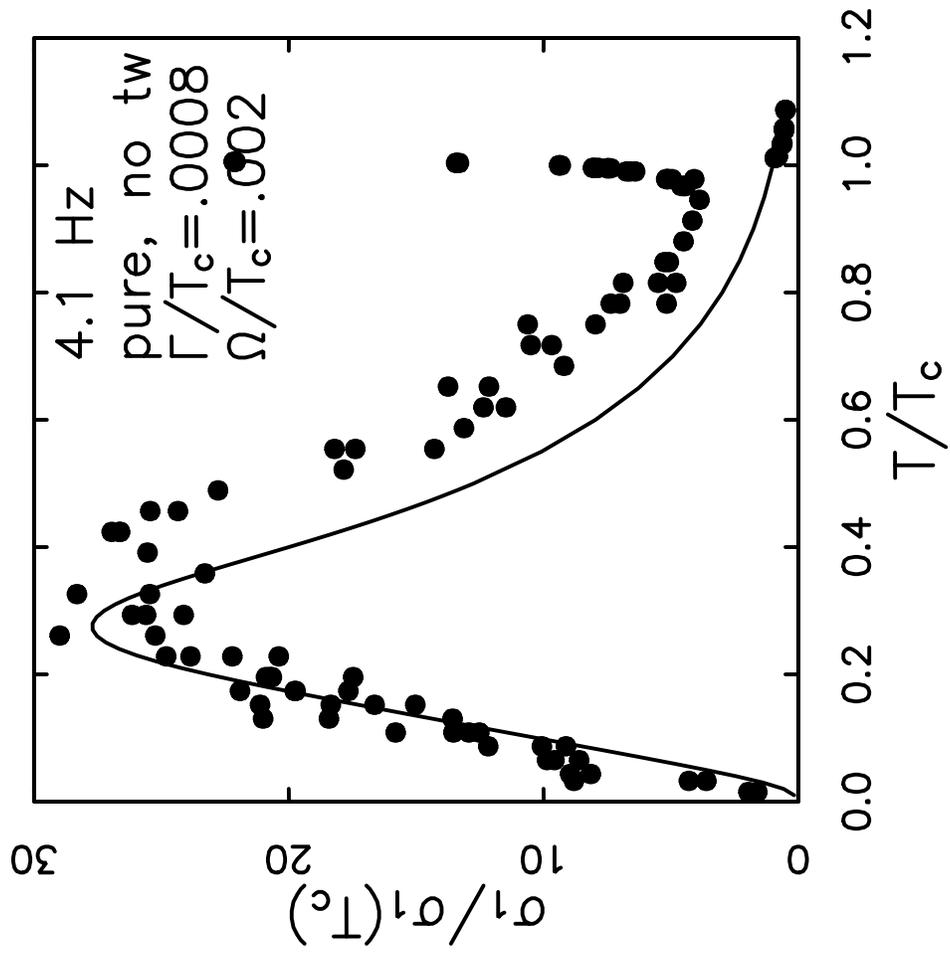
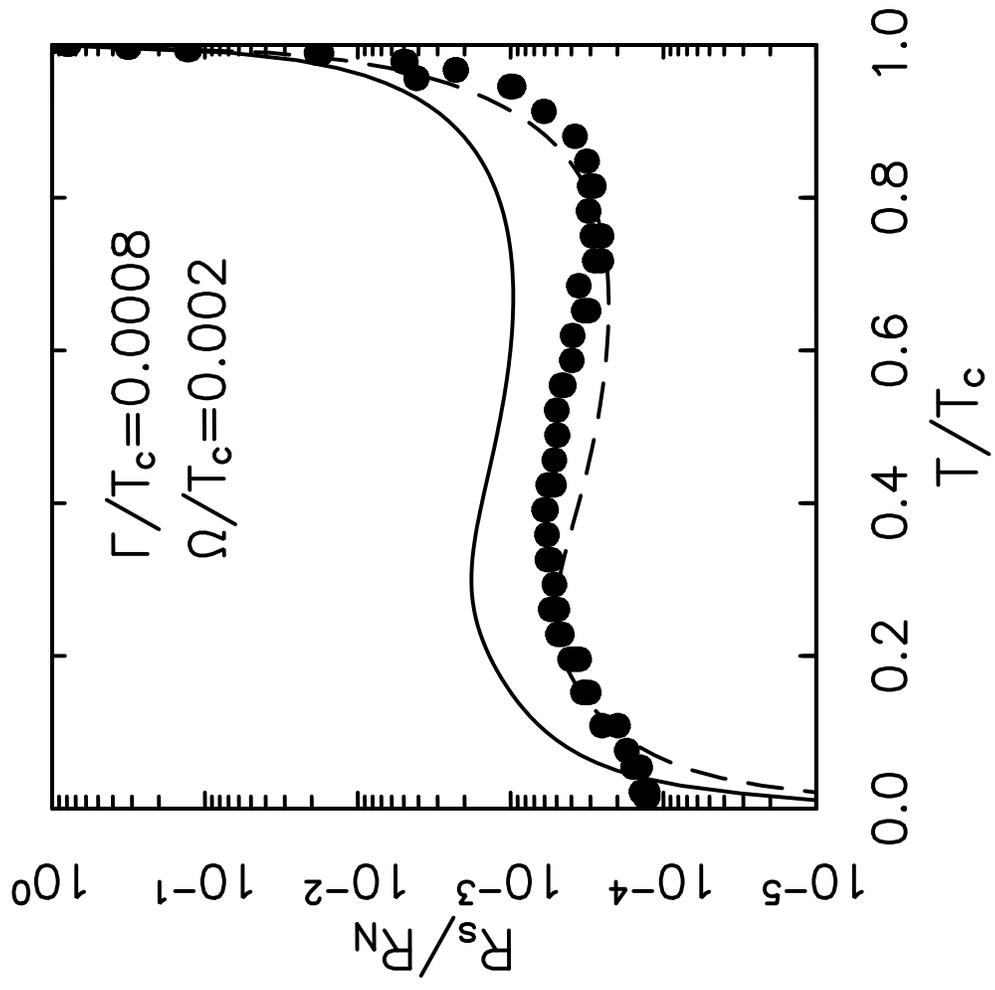
\end{document}

***************************************

**************   FIGURES  *************

***************************************

FIGURE 1